\documentclass[11pt,a4paper]{article}
\pdfoutput=1

\usepackage{a4}
\usepackage{jheppub}
\usepackage{graphics}
\usepackage{amssymb,amsmath}
\usepackage{color}
\usepackage{wasysym}
\usepackage[tight]{subfigure}
\usepackage{booktabs}
\usepackage{rotating,multirow,booktabs}

\newcommand{\beq}{\begin{equation}}
\newcommand{\eeq}{\end{equation}}
\newcommand{\beqn}{\begin{eqnarray}}
\newcommand{\eeqn}{\end{eqnarray}}
\newcommand{\bmat}{\begin{pmatrix}}
\newcommand{\emat}{\end{pmatrix}}

\def\rtwo{\sqrt{2}}

\def\nthree{\tilde \chi_3^0}
\def\nfour{\tilde \chi_4^0}

\def\chapmone{\tilde \chi_1^\pm}

\def\chapmtwo{\tilde \chi_2^\pm}
\def\twz{\tilde W^0}
\def\twpm{\tilde W^\pm}
\def\twmp{\tilde W^\mp}

\def\om{\omega}

\def\lsim{\raise0.3ex\hbox{$\;<$\kern-0.75em\raise-1.1ex\hbox{$\sim\;$}}}
\def\gsim{\raise0.3ex\hbox{$\;>$\kern-0.75em\raise-1.1ex\hbox{$\sim\;$}}}

\newcommand{\gev}{\ \mathrm{GeV}}
\newcommand{\tev}{\ \mathrm{TeV}}

\def\gsim  {\hspace{0.3em}\raisebox{0.4ex}{$>$}\hspace{-0.75em}\raisebox{-.7ex}{$\sim$}\hspace{0.3em}}
\def\lsim  {\hspace{0.3em}\raisebox{0.4ex}{$<$}\hspace{-0.75em}\raisebox{-.7ex}{$\sim$}\hspace{0.3em}}

\def\cb{c_\beta}
\def\sb{s_\beta}
\def\ctb{c_{2\beta}}
\def\stb{s_{2\beta}}
\def\cw{c_W}
\def\sw{s_W}

\def\stw{s_{2W}}

\preprint{\tiny KCL-PH-TH/2015-30, LCTS/2015-22, IFT-UAM/CSIC-15-070 \normalsize}
\title{ Long-lived bino and wino in supersymmetry with heavy scalars and higgsinos }
\author[a]{Krzysztof Rolbiecki}
\author[b]{and Kazuki Sakurai}
\affiliation[a]{
Instituto de F\'{\i}sica Te\'{o}rica, IFT-UAM/CSIC,\\
C/ Nicol\'{a}s Cabrera, 13-15, Cantoblanco, 28049 Madrid, Spain}
\affiliation[b]{Department of Physics, King's College London, London WC2R 2LS, UK}
\emailAdd{rolbiecki.krzysztof@csic.es}
\emailAdd{kazuki.sakurai@kcl.ac.uk}
\abstract{
We point out that there is a parameter region in supersymmetry with heavy scalars and higgsinos, in which
the heavier of bino and wino 
becomes long-lived as a consequence of the heavy higgsinos.  
In this region these electroweak gaugino sectors 
are secluded from each other with very small mixings 
that are inversely proportional to the higgsino mass, $\mu$.
We revisit the bino and bino decays and provide simple formulae for the partial decay rates 
and the lifetimes in the limit of heavy higgsinos.
The scale of $\mu$ required for the long-lived electroweak gauginos
highly depends on the mass hierarchy between bino and wino.  The neutral wino
can be long-lived ($c \tau \gsim 1$~cm) even with $|\mu| \gsim 10$~TeV
if $ m_{\tilde W} - m_{\tilde B} \sim 20$~GeV.  
We discuss the collider signatures of the long-lived binos and winos in this scenario.
}

\begin{document}
\maketitle
\flushbottom


\section{Introduction\label{intro}}

Supersymmetry (SUSY) remains as a promising new physics candidate after LHC Run-I.
The discovery of the Higgs boson with $m_h \simeq$ 125 GeV~\cite{Aad:2012tfa,Chatrchyan:2012ufa,Khachatryan:2014jba,Aad:2015zhl} and the negative results of the new physics searches at the LHC make the SUSY scenario with heavy scalars more attractive.
In such a scenario, scalars (except for the Standard Model (SM) like Higgs boson, $h$) are heavier than the TeV scale and the observed Higgs mass can be easily realised by the large corrections from the heavy scalars~\cite{Giudice:2011cg, Ibe:2012hu, Bagnaschi:2014rsa}.
The gauge coupling unification can be achieved with the light gauginos~\cite{ArkaniHamed:2004fb} and such light gauginos can be within the reach of the collider experiments. 
If the wino is the lightest SUSY particle (LSP), the thermal relic abundance bounds the wino mass from above by $2.7-3$ TeV
\cite{Hisano:2006nn, Cirelli:2007xd, Cohen:2013ama}.
Phenomenology of light neutralinos with the thermal relic density smaller than the observed dark matter abundance was extensively studied in the literature \cite{Cheung:2012qy, Berggren:2013vfa, Han:2013kza, Schwaller:2013baa, Low:2014cba, Anandakrishnan:2014exa, Bramante:2014dza, Martin:2014qra, Acharya:2014pua, diCortona:2014yua, Han:2015lma, Barducci:2015ffa, Badziak:2015qca}.
The splitting between the scalar and gaugino masses is also motivated from model building perspective.
If the SUSY breaking field with non-vanishing $F$-term is charged under some symmetry, only scalars acquire the soft masses at tree-level.
The gaugino masses can be generated by other mechanisms such as the anomaly mediation, but they are parametrically smaller than the scalar masses.
  
It has been pointed out that gluinos can be long-lived in this scenario if squarks are heavier than ${\cal O} (10^3)$ TeV~\cite{ArkaniHamed:2004fb}.  
Such long-lived gluinos, if produced, may leave a distinctive $R$-hadron signature or a signature of displaced vertices~\cite{Hewett:2004nw, Aad:2013gva}.
With decoupled squarks, the $SU(3)$ gaugino sector is secluded from the other gaugino sectors 
and the gluino becomes stable even if there exist lighter gauginos.
The situation is somewhat different for the $SU(2)_L$ and $U(1)_Y$ gauginos 
because of the electroweak (EW) symmetry breaking (EWSB).
These sectors are mixed after the EWSB and 
the heavier gaugino can decay into lighter ones without mediation of scalars.
However, this mixing is proportional to the ratios of the vacuum expectation values of the Higgs field and the mass of the fermionic component of the Higgs field: $v_u/\mu$ and $v_d/\mu$,  
and can become arbitrarily small if the magnitude of $\mu$ increases.

In this paper we point out that there is a region in SUSY parameter space with heavy scalars and higgsinos where the heavier of  EW gauginos becomes long-lived as a consequence of the large $\mu$ parameter.
We revisit the decays of EW gauginos and derive simple formulae for partial decay rates and lifetimes,
which are valid in the limit of the heavy higgsinos.
We identify the region where the bino and winos become long-lived 
for both cases of two- and three-body decays.

The paper is organised as follows. In the following section we review the chargino and neutralino sectors
of the minimal supersymmetric Standard Model (MSSM)
and study the mixings and interactions between bino and winos in the heavy higgsino limit.
We then revisit the two- and three-body decays of bino and winos 
in sections \ref{sec:2body} and \ref{sec:3body}, respectively.
In these sections, we identify the scale of $\mu$, at which the bino and winos become long-lived in collider experiments.
In section~\ref{sec:splitting} we briefly discuss how the large mass splitting between the gauginos and higgsinos 
can be theoretically achieved.
The collider signatures of the long-lived bino and winos in our scenario is briefly discussed in section~\ref{sec:collider}.
Finally, we conclude in section~\ref{sec:conclusion}. In appendix~\ref{app:funcs} we list detailed formulae for auxiliary functions used in the paper and in appendix~\ref{sec:comparison} we compare approximate expressions obtained here with an exact calculation using \texttt{SDecay}~\cite{Muhlleitner:2003vg}.


\section{Interactions between winos and bino\label{sec:interaction}}

The chargino mass matrix is given by:\footnote{
We closely follow the calculation and conventions used in
\cite{Gunion:1992tq, Djouadi:2001fa}.
}
\beq
{\cal M}_C =
\bmat
M_2 & \sqrt{2} m_W \sb \\
\sqrt{2} m_W \cb & \mu
\emat,
\eeq
where $c_\beta$ ($s_\beta$) represents $\cos \beta$ ($\sin \beta$) and $\beta$ is defined as the ratio of the vacuum expectation values of two Higgs fields,
$\tan\beta = \langle H_u \rangle / \langle H_d \rangle$.
In the large $\mu$ limit, this matrix can be diagonalised as ${\cal M}_C^{\rm diag} = U^* {\cal M}_C V^{-1}$ where $U$ and $V$ take the following forms
\beq
U \simeq
\bmat
1 & - \frac{\sqrt{2} \sb m_W}{\mu} \\
\frac{\sqrt{2} \sb m_W}{\mu} & 1 
\emat, ~~~~~~~
V \simeq
\bmat
1 & - \frac{\sqrt{2} \cb m_W}{\mu} \\
\frac{\sqrt{2} \cb m_W}{\mu} & 1 
\emat.
\eeq
The eigenvalues are then given by: 
\beq
m_{\tilde W^{\pm}} \simeq M_2,~~~~ m_{\chapmtwo} \simeq |\mu| ,
\eeq
with $m_{\twpm} \ll m_{\chapmtwo}$.\footnote{In order to stress wino nature of the lightest chargino, we denote it here as $\tilde W^{\pm}$ rather then usual $\chapmone$.}

The neutralino mass matrix is given by
\beq
{\cal M}_N =
\bmat
M_1 & 0 & -m_Z s_W \cb & m_Z s_W \sb \\
0 & M_2 & m_Z c_W \cb & -m_Z c_W s_\beta \\
-m_Z s_W \cb & m_Z c_W \cb & 0 & -\mu \\
m_Z s_W \sb & -m_Z c_W \sb & -\mu & 0  
\emat.
\eeq
In the large $\mu$ limit, the diagonalising matrix $Z$ can be written as 
\beq \label{eq:zapprox}
Z \simeq
\bmat
1 & -\frac{m_Z^2 \stw \stb}{2 \mu (M_2 - M_1)} & \frac{m_Z \sw \sb}{\mu}  & - \frac{m_Z \sw \cb}{\mu}  \\
\frac{m_Z^2 \stw \stb}{2 \mu (M_2 - M_1)} & 1 & - \frac{m_Z \cw \sb}{\mu}  & \frac{m_Z \cw \cb}{\mu}  \\
\frac{m_Z \sw (\cb - \sb)}{\sqrt{2} \mu} & -\frac{m_Z \cw (\cb - \sb)}{\sqrt{2} \mu} & \frac{1}{\rtwo} & \frac{1}{\rtwo} \\
\frac{m_Z \sw (\cb + \sb)}{\sqrt{2} \mu} & -\frac{m_Z \cw (\cb + \sb)}{\sqrt{2} \mu} & -\frac{1}{\rtwo} & \frac{1}{\rtwo} 
\emat,
\eeq
where $s_W$ is the sine of Weinberg angle, $\sin \theta_W$, and $c_W \equiv \cos \theta_W$.  
The mass eigenvalues are found to be
\beq
m_{\tilde B} \simeq M_1,~~~m_{\tilde W^0} \simeq M_2,~~~m_{\tilde \chi_3^0} \simeq m_{\tilde \chi_4^0} \simeq |\mu|\,,
\eeq
where $m_{\tilde B}, m_{\tilde W^0} \ll m_{\nthree}, m_{\nfour}$.

\begin{figure}[t!]
\begin{center}
\includegraphics[width=0.28\textwidth]{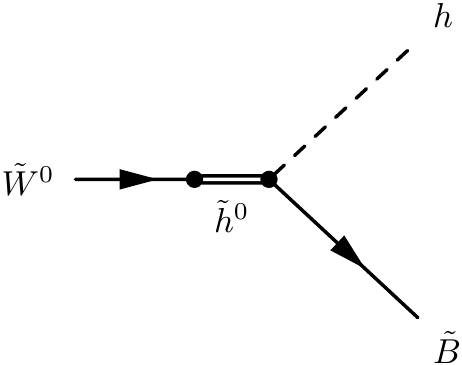}\label{fig:W-h_1}
\hspace{2mm}
\includegraphics[width=0.28\textwidth]{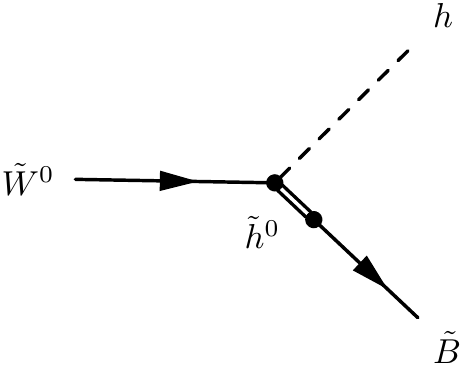}\label{fig:W-h_2}
\\
$~~~~~~~~~~~\propto 1/\mu$
\end{center}
\begin{tabular}{cc}
\begin{minipage}{0.5\hsize}
\begin{center}
\includegraphics[width=0.6\textwidth]{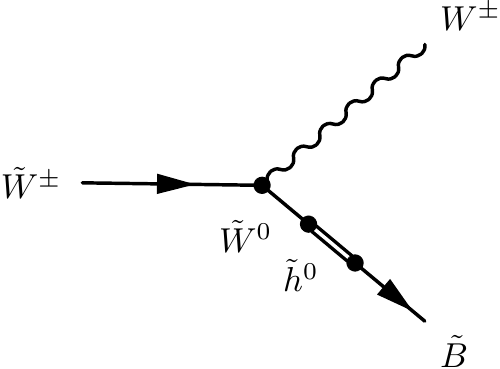}\label{fig:Wpm-wpm}
\\
$\propto 1/\mu$
\end{center}
\end{minipage}
\begin{minipage}{0.5\hsize}
\begin{center}
\includegraphics[width=0.6\textwidth]{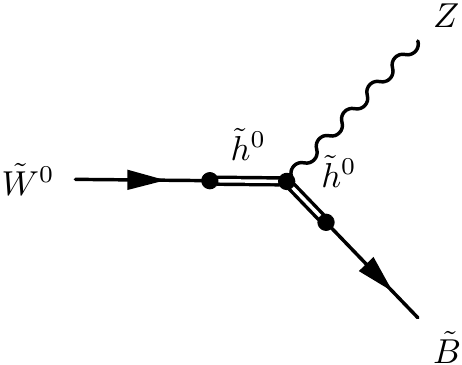}\label{fig:W-z}
\\
$\propto 1/\mu^2$
\end{center}
\end{minipage}
\end{tabular}
\caption{The wino-bino interaction with the SM Higgs and gauge bosons.}
\label{fig:interaction}
\end{figure} 

The wino and bino interact with each other and a Higgs or a gauge boson through the mixing matrices $U$, $V$ and $Z$.
The interaction with the light Higgs boson, $h$, is dictated by the $\tilde \chi^0_j$-$\tilde \chi_i^0$-$h$ coupling~\cite{Djouadi:2001fa} 
\beqn
G_{ijh}^{L} = G_{ijh}^{R} = - \frac{1}{2 \sw} (Z_{j2} - Z_{j1} \tan \theta_W )(Z_{i3} \sin \alpha  + Z_{i4} \cos \alpha ) + i \leftrightarrow j\,,
\eeqn
where $\alpha$ is the mixing angle of the neutral Higgs sector and reduces to 
$\sin \alpha = \cb$, $\cos \alpha = -\sb$ in the decoupling limit of the SUSY Higgses~\cite{Haber:1995be,Djouadi:1996pj,Djouadi:2005gj}.
In the large $\mu$ limit these couplings can be written as
\beqn
G_{\twz \tilde B h} \simeq
\frac{1}{2 \sw} \Big[ (\cb Z_{13} - \sb Z_{14}) - \frac{\sw}{\cw} (\cb Z_{23} - \sb Z_{24})
\Big]
\simeq
\frac{m_Z \stb}{\mu}\,.
\label{GWBh}
\eeqn
This interaction is illustrated diagrammatically in the top-row graphs of figure~\ref{fig:interaction}.   
We note that the coupling originates from the bino-higgsino and wino-higgsino mixing and is suppressed by $m_Z/\mu$ for large $\mu$.

The interaction with the $W$ boson is described by the $\tilde \chi^\pm_j$-$\tilde \chi_i^0$-$W^\mp$ coupling 
\beqn
G_{ijW}^L = \frac{1}{\sw}( Z_{i2} V_{j1} - \frac{1}{\rtwo} Z_{i4} V_{j2})\,, ~~~
G_{ijW}^R = \frac{1}{\sw}( Z_{i2} U_{j1} + \frac{1}{\rtwo} Z_{i3} U_{j2})\,, 
\eeqn
and in the large $\mu$ limit these couplings can be written as 
\beq
G_{\tilde B  \twpm  W} \simeq  -\frac{m_Z^2 \cw \stb}{\mu (M_2 - M_1)}\, .
\eeq  
The interaction is depicted in the bottom-left graph of figure~\ref{fig:interaction}.   
The coupling is proportional to $m_Z/\mu$ and originates from the bino-wino mixing, $Z_{12}$. 
  
We illustrate the wino-bino interaction with the $Z$ boson in the bottom-right graph of figure~\ref{fig:interaction}.   
Unlike the other interactions, this coupling requires both the wino-higgsino and the bino-higgsino interaction.
Formally, the interaction is defined by the $\tilde \chi^0_j$-$\tilde \chi_i^0$-$Z$ coupling 
\beq \label{eq:GijZ}
G_{ijZ}^R = - G_{ijZ}^L = \frac{1}{\stw}( Z_{i3} Z_{j3} - Z_{i4} Z_{j4})\,,
\eeq
and reduces in the large $\mu$ limit to 
\beq
G_{\twz \tilde B Z}^R = - G_{\twz \tilde B Z}^L  \simeq  \frac{m_Z^2 \ctb}{2 \mu^2} \,.
\label{eq:GWBZ}
\eeq
These couplings are proportional to $(m_Z/\mu)^2$ and become small 
quicker than the other couplings as $|\mu|$ increases.    
There exists another contribution to the bino-wino-$Z$ interaction from higher order terms. 
The higgsino-Higgs loop diagram generates a dimension-5 operator,
\beq
\tilde B \tilde W^a   \sigma^{\mu \nu} F^a_{\mu \nu}\,.
\label{eq:dim5}
\eeq
A naive dimensional analysis suggests that the coefficient of this operator is proportional to
$\alpha / (4 \pi \mu)$.
Although this contribution is suppressed by the loop factor, 
it can easily dominate the tree-level interaction, eq.~(\ref{eq:GWBZ}), for very large $\mu$.
We leave a detailed study of effects of this operator for future work.

\section{Two-body decays}\label{sec:2body}   

\subsection{Wino NLSP case ($|M_2| > |M_1|$)}\label{sec:3.1}

Throughout this paper, we assume that gluinos are heavier than wino and bino.
In this section we consider the cases where winos are heavier than bino.
In SUSY models with heavy scalars and higgsinos,
the mass difference between the charged and neutral winos is small
and the decays among the wino multiplet (e.g.\ $\twpm \to \pi^\pm \twz$) can be neglected 
compared to the wino decays into a bino.
We therefore do not consider the decays within the wino multiplet in this section.
We will mention the effect of this decay mode in section~\ref{sec:collider}.

The decay mode of the charged wino is shown in the bottom-left diagram of figure~\ref{fig:interaction}.
The final state consists of $W^\pm$ and $\tilde B$.
The decay rate of this process is given by~\cite{Djouadi:2001fa}:
\beqn
\Gamma( \twpm \to W^\pm \tilde B) 
&=& \frac{\alpha}{4} M_2 G_{\tilde B  \twpm  W}^2 f_- (\mu_{\tilde B}, \mu_W) 
\nonumber \\ 
&\stackrel{\mu \gg m_Z}{\simeq}& 
 \frac{\alpha \stb^2 M_2}{4} \frac{m_Z^2}{\mu^2} \frac{m_W^2 f_- (\mu_{\tilde B}, \mu_W)}{(M_2 - M_1)^2}  
\nonumber \\ 
&\stackrel{M_2 \gg m_W}{\simeq}& 
\frac{\alpha}{4} (M_2 - M_1) \stb^2 \frac{m_Z^2}{\mu^2} \left(1 + \frac{M_1}{M_2} \right)^3 \,,
\label{eq:Wpm_decay}
\eeqn
where $\mu_{\tilde B} = M_1^2/M_2^2$ and $\mu_W = m^2_W/M_2^2$.
The analytic expression for $f_-(x,y)$ is given in appendix~\ref{app:funcs}.
Throughout this and the following sections, the first approximation is valid in the large $\mu$ limit, while the 
following assumes both $\mu \gg m_Z$ and $M_2 \gg m_W$. 
Using the last expression, the lifetime of the charged wino is found to be
\beq
c \tau_{\tilde W^\pm} \simeq 2.5 \,{\rm cm} \cdot 
\left( \frac{\mu}{10^6\,\tev} \right)^{2}
\left( \frac{500\,\gev}{M_2 - M_1} \right)
\left( \frac{1}{s_{2\beta}} \right)^{2}
\left( \frac{M_2}{ M_2 + M_1 } \right)^3 \,.
\label{eq:lifetime_W2d}
\eeq

\begin{figure}[t!]
\begin{center}
  \subfigure[]{\includegraphics[width=0.5\textwidth]{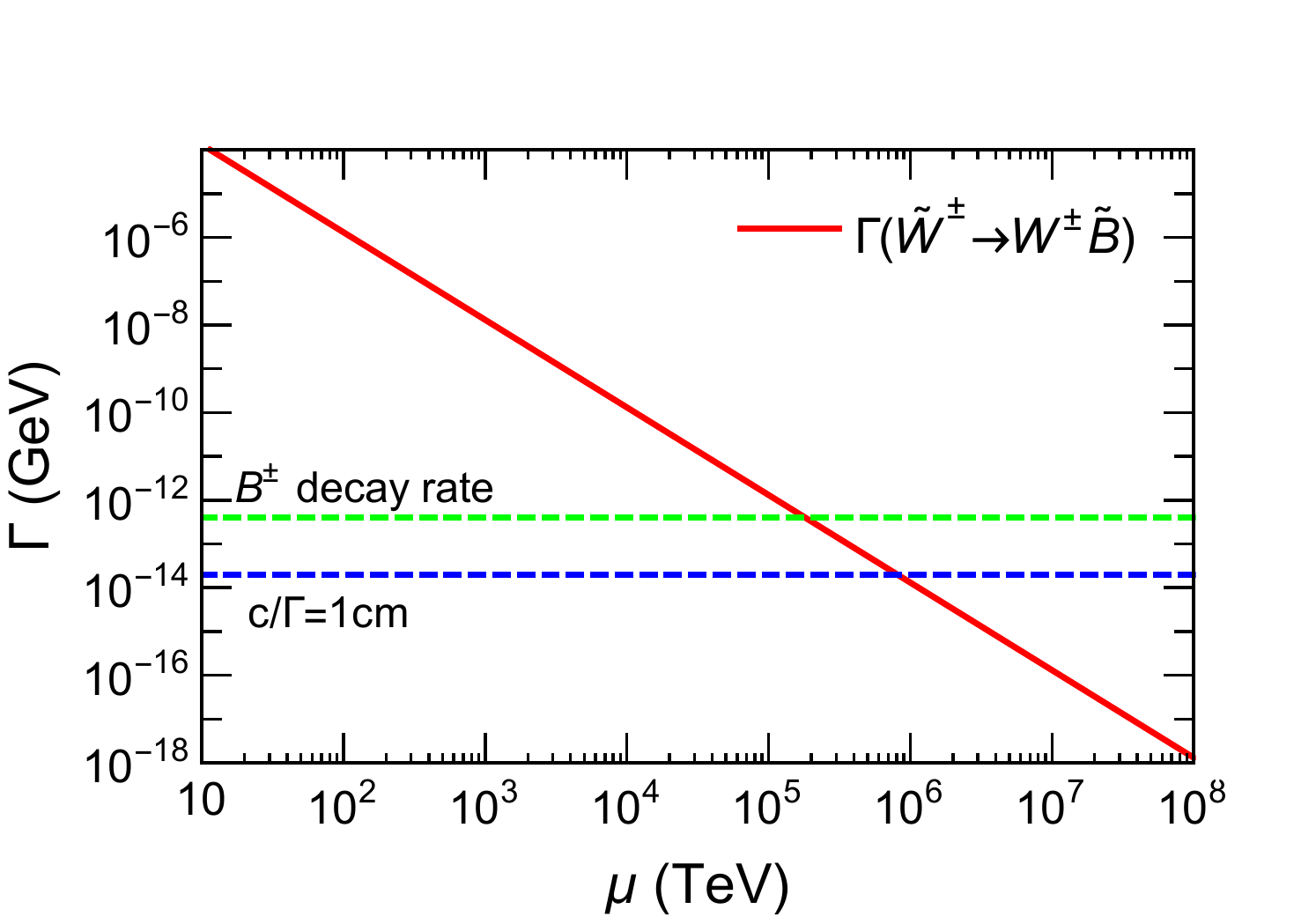}\label{fig:GamMu_Wpm-wpm}} \hspace{-4mm}
  \subfigure[]{\includegraphics[width=0.5\textwidth]{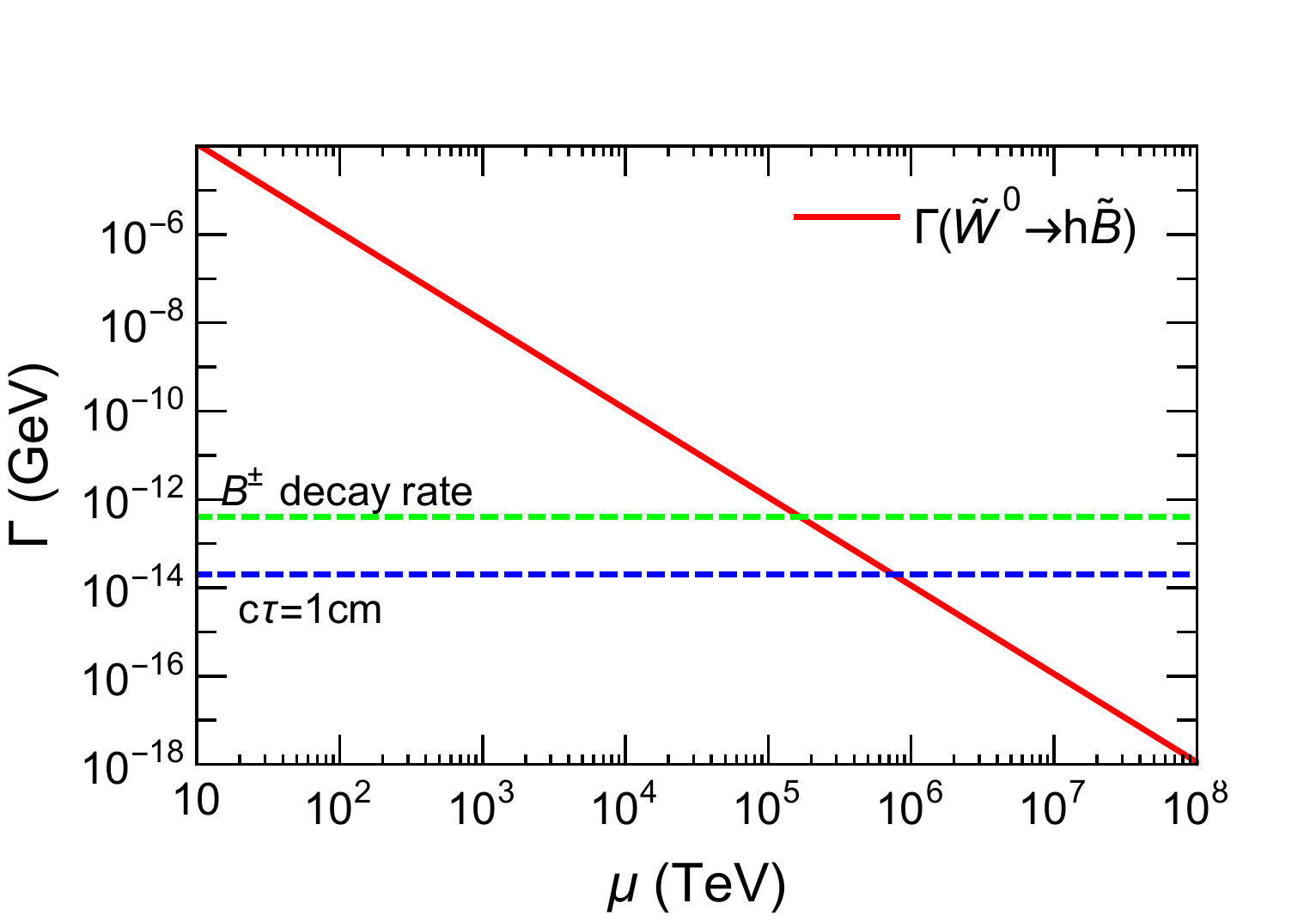}\label{fig:GamMu_W0-zhB}} 
\caption{ The two-body decay rate of winos as a function of $\mu$, with other parameters fixed at:  $M_2 = 800$ GeV, $M_1 = 300$ GeV, $\tan \beta = 2$.   
\label{fig:GamMu_2B} }
\end{center}
\end{figure}

In figure~\ref{fig:GamMu_Wpm-wpm} we show the decay rate of $\twpm$ as a function of $|\mu|$.
The horizontal green line denotes the decay rate of $B^\pm$ meson as a reference,
where the displacement of the $\twpm$ decay from the primary vertex starts to be visible.
The horizontal blue line corresponds to the decay rate where the $c / \Gamma = 1$~cm,
where the charged winos start to reach the trackers leaving the distinctive kink-like signature.\footnote{The trackers inner radii are for example: 44~mm for CMS~\cite{Chatrchyan:2014fea}, 50.5~mm for ATLAS~\cite{ATLAS-CONF-2014-047}, and 16~mm for ILD at the planned International Linear Collider~\cite{Behnke:2013lya}.} 
As can be seen, the charged wino has a collider relevant lifetime for $|\mu| \gsim {\cal O}(10^5)$ TeV,
if the two body $\twpm \to W^\pm \tilde B$ is kinematically allowed.

There are two possible decay modes for the neutral wino:
$\twz \to Z \tilde B$ and $\twz \to h \tilde B$.
The former is depicted in the bottom-right diagram in figure~\ref{fig:interaction}
and can be mediated either by the coupling eq.~\eqref{eq:GijZ} or by the dimension-5 operator, eq.~\eqref{eq:dim5}.  
However, these contributions are suppressed by the extra $m_Z/\mu$ factor and the loop factor in the matrix element, respectively, 
compared to the $\twz \to h \tilde B$ decay.
Therefore, if the mass difference is large enough to allow the two body $\twz \to h \tilde B$ decay,  
the neutral wino predominantly decays into $h$ and $\tilde B$.
The decay rate is given by: 
\beqn
\Gamma( \twz \to h \tilde B) 
&=& 
\frac{\alpha}{4} M_2 G_{\twz  \tilde B  h}^2 f_h (\mu_{\tilde B}, \mu_h)
\nonumber \\ 
&\stackrel{\mu \gg m_Z}{\simeq}& 
\frac{\alpha  M_2}{4} \stb^2 \frac{m_Z^2}{\mu^2} f_h (\mu_{\tilde B}, \mu_h)
\nonumber \\ 
&\stackrel{M_2 \gg m_h}{\simeq}&  
\frac{\alpha }{4} ( M_2 - M_1) \stb^2 \frac{m_Z^2}{\mu^2}  \left( 1 + \frac{M_1}{M_2} \right)^3\,,
\eeqn
where $\mu_h = m^2_h/M_2^2$, and the analytic expression of $f_h(x,y)$ is given in appendix~\ref{app:funcs}.
We approximate the exact formula for $\mu \gg m_Z$ and, additionally, $M_2 \gg m_h$ in the second and third line, respectively.
It is worth noting that the last expression is identical to the one in eq.~\eqref{eq:Wpm_decay}.
This can be also seen in figure~\ref{fig:GamMu_W0-zhB},
where we show the decay rate of $\twz$ as a function of $|\mu|$.
The approximate lifetime formula for the $\twz$ is therefore the same as in eq.~\eqref{eq:lifetime_W2d} with $c \tau_{\tilde W^0} \sim c \tau_{\tilde W^\pm}$. 
At colliders, the long-lived neutral wino leaves the displaced dijet/jets signatures~\cite{CMS:2014wda,Aad:2015asa}.

Finally, we would like to comment on a special case when $m_Z < |M_2| - |M_1| < m_h$.
In this case the $\twz \to h \tilde B$ decay is kinematically forbidden and the $\twz$ decays predominantly into $Z$ and $\tilde B$,
through the bottom-right diagram in figure~\ref{fig:interaction} and the dimension-5 operator, eq.~\eqref{eq:dim5}.
Since the tree-level contribution is suppressed by the extra $1/\mu$ factor, the contribution from the dimension-5 operator 
dominates in the large $|\mu|$ region.   
This parameter region is interesting, because
the neutral wino can be long lived with much smaller $|\mu|$.
We leave this question for future work.

\subsection{Bino NLSP case ($|M_1| > |M_2|$)}

If bino is heavier than winos, the bino can decay either to $\tilde B \to \twpm W^\mp$, or $ \tilde B \to h \tilde W^0$, or $ \tilde B \to Z \tilde W^0$.
As we discussed in the previous subsection, the interaction $\tilde B$-$\tilde W$-$Z$ is suppressed compared to $\tilde B$-$\tilde W$-$W$ and $\tilde B$-$\tilde W$-$h$,
and the bino decay is dominated by the $W$ and $h$ decay modes.
The partial decay rates for these modes are given by:
\beqn
\Gamma( \tilde B \to \twpm W^\mp) 
&=& 
\frac{\alpha}{2} M_1 G_{\tilde B  \twpm  W}^2 f_- (\mu_{\tilde W}, \mu_W)
\nonumber \\ 
&\stackrel{\mu \gg m_Z}{\simeq}&
\frac{\alpha \stb^2 M_1}{2} \frac{m_Z^2}{\mu^2}  \frac{m_W^2 f_- (\mu_{\tilde W}, \mu_W)}{(M_1 - M_2)^2} 
\nonumber \\ 
&\stackrel{M_1 \gg m_W}{\simeq}&
\frac{\alpha}{2} (M_1 - M_2) \stb^2 \frac{m_Z^2}{\mu^2} \left(1 + \frac{M_2}{M_1} \right)^3 \,,
\eeqn
and
\beqn
\Gamma( \tilde B \to h \tilde W^0) 
&=& 
\frac{\alpha}{4} M_1 G_{\twz  \tilde B  h}^2 f_h (\mu_{\tilde W}, \mu_h) 
\nonumber \\
&\stackrel{\mu \gg m_Z}{\simeq}&
\frac{\alpha  M_1}{4} \stb^2 \frac{m_Z^2}{\mu^2} f_h (\mu_{\tilde W}, \mu_h)
\nonumber \\
&\stackrel{M_1 \gg m_h}{\simeq}&
\frac{\alpha}{4} (M_1 - M_2) \stb^2 \frac{m_Z^2}{\mu^2} \left(1 + \frac{M_2}{M_1} \right)^3 \,,
\eeqn
where $\mu_{\tilde W} = M_2^2/M_1^2$, $\mu_W = m^2_W/M_1^2$ and $\mu_h = m^2_h/M_1^2$. 
The approximate formulae are for successively applied $M_1 \gg m_W$ and $M_1 \gg m_h$, respectively.  
If $M_1 \gg m_h$, we find that the branching ratios for different decay modes are approximately equal: 
\beq
BR(\tilde B \to W^+ \tilde W^-) = BR(\tilde B \to W^- \tilde W^+) \sim BR(\tilde B \to h \tilde W^0) \sim 1/3.
\eeq

\begin{figure}[t!]
\begin{center}
  \includegraphics[width=0.6\textwidth]{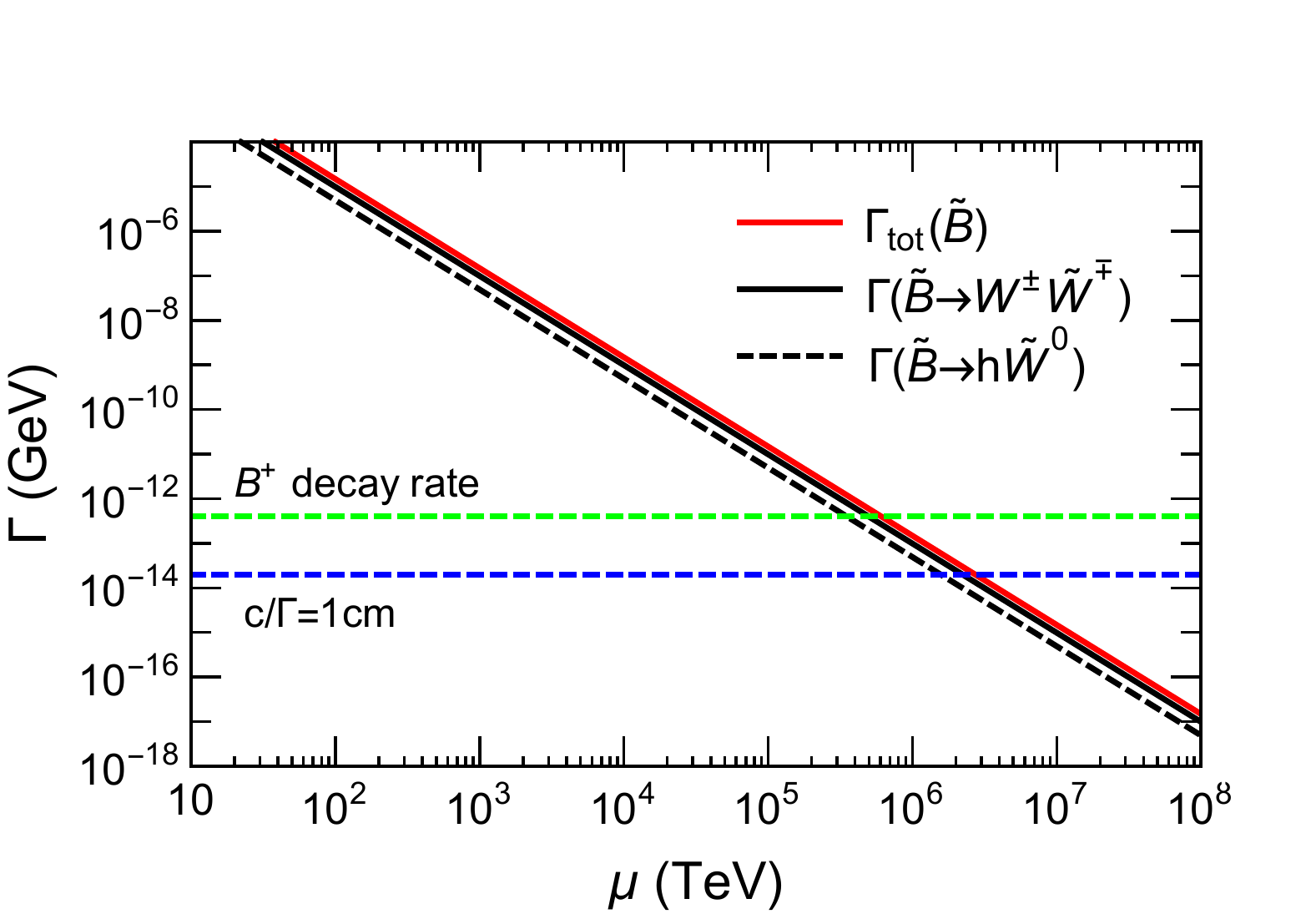}
\caption{
The two-body decay rate of bino as a function of $\mu$, with other parameters fixed at:  $M_1 = 800$ GeV, $M_2 = 300$ GeV, $\tan \beta = 2$.   
\label{fig:GamMu_B-whW} }
\end{center}
\end{figure}

Figure~\ref{fig:GamMu_B-whW} shows the total and partial decay rates for the bino as a function of $|\mu|$.
One can see that the bino can be long-lived on collider time-scales if $|\mu| \gsim {\cal O}(10^6)$~TeV.
The bino lifetime is approximately given as
\beq
c \tau_{\tilde B} \simeq 1 \,{\rm cm} \cdot 
\Big( \frac{\mu}{10^6\,\tev} \Big)^{2}
\Big( \frac{500\,\gev}{M_1 - M_2} \Big)
\Big( \frac{1}{s_{2\beta}} \Big)^{2}
\left( \frac{M_1}{ M_1 + M_2 } \right)^3~.
\eeq


\section{Three-body decays}\label{sec:3body}

In this section we consider the cases where the mass difference between bino and wino is small and two-body decay modes
considered in the previous section are kinematically not allowed.

\subsection{Wino NLSP case ($|M_2| > |M_1|$)}\label{sec:4.1}

\begin{figure}[t!]
\begin{center}
\includegraphics[width=0.32\textwidth]{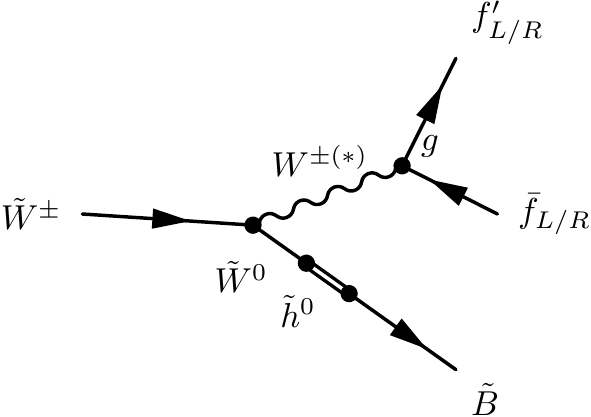}\label{fig:W-w-ffB}
\\$\propto g/\mu$
\end{center}
\caption{Diagram for $\twpm \to f' \bar{f} \tilde B$ three-body decay.}
\label{fig:Wpm_offshellW}
\end{figure} 

We start with a case of $|M_2| > |M_1|$ and define the mass difference $\Delta M \equiv |M_2| - |M_1| > 0$. 
If $\Delta M < m_W$ the charged wino two-body decay, $\twpm \to W^\pm \tilde B$, is forbidden.
In this case, the charged wino decays into a pair of fermions and a bino, via an off-shell $W$ as shown in figure~\ref{fig:Wpm_offshellW}. 
The decay rate of this process is given by:
\beqn
\Gamma_f(\tilde W^+ \to f \bar f' \tilde B) 
&=& 
\frac{\alpha^2 M_2 }{16 \pi} \frac{1}{\sw^2}   \cdot G_{\tilde B \tilde W^\pm W}^2 \cdot \Omega_- (\mu_{\tilde B}, \mu_W)
\nonumber \\
&\stackrel{\mu \gg m_Z}{\simeq}&
\frac{\alpha^2 M_2 }{16 \pi} \frac{\stb^2}{\sw^2} \frac{m_Z^2}{\mu^2} 
\frac{m_W^2 \Omega_-(\mu_{\tilde B}, \mu_W)}{(M_2 - M_1)^2} 
\nonumber \\
&\stackrel{\Delta M \ll M_2}{\simeq}&
\frac{4 \alpha^2 }{15 \pi} \frac{ (\Delta M)^3 }{\mu^2} \frac{\stb^2}{\stw^2} \,,
\eeqn
where $\mu_W = m_W^2/M_2^2$ and an analytic expression of $\Omega_\pm(x,y)$ is given in appendix~\ref{app:funcs}, eqs.~\eqref{eq:omegaapp}--\eqref{eq:ellxy}.
The last approximation assumes $\Delta M \ll M_2$ together with $\mu \gg m_Z$.  
One may naively expect that the decay rate should be proportional to $(\Delta M)^5 / m_W^4$ due to the $s$-channel structure.
However the wino-bino mixing, $Z_{12}$, is inversely proportional to $\Delta M$, see eq.~\eqref{eq:zapprox}, and
the decay rate is finally proportional to $(\Delta M)^3$. 

\begin{figure}[t!]
\begin{center}
  \subfigure[]{\includegraphics[width=0.5\textwidth]{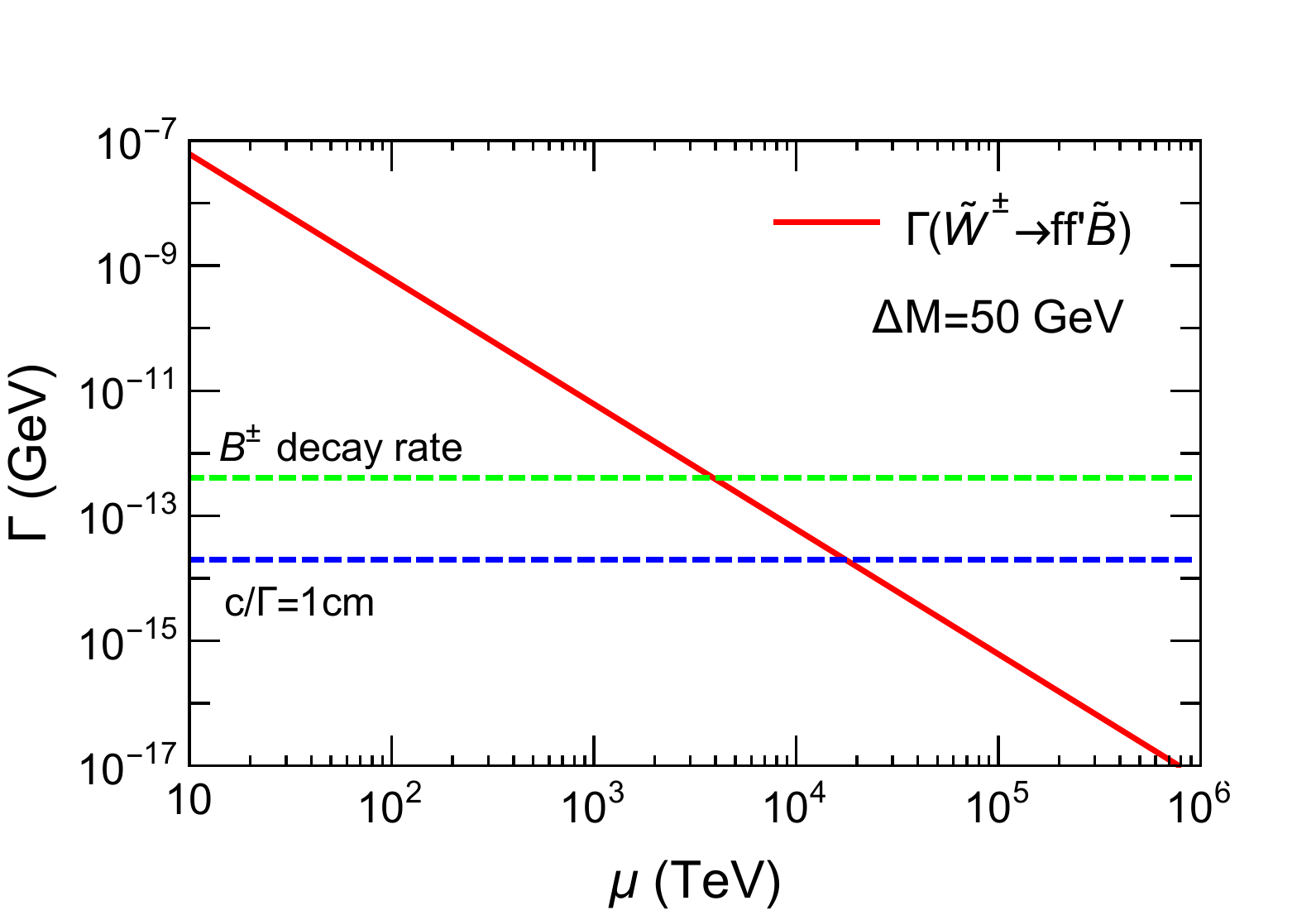}\label{fig:GamMu_Wpm-ffB_50}} \hspace{-4mm}
  \subfigure[]{\includegraphics[width=0.5\textwidth]{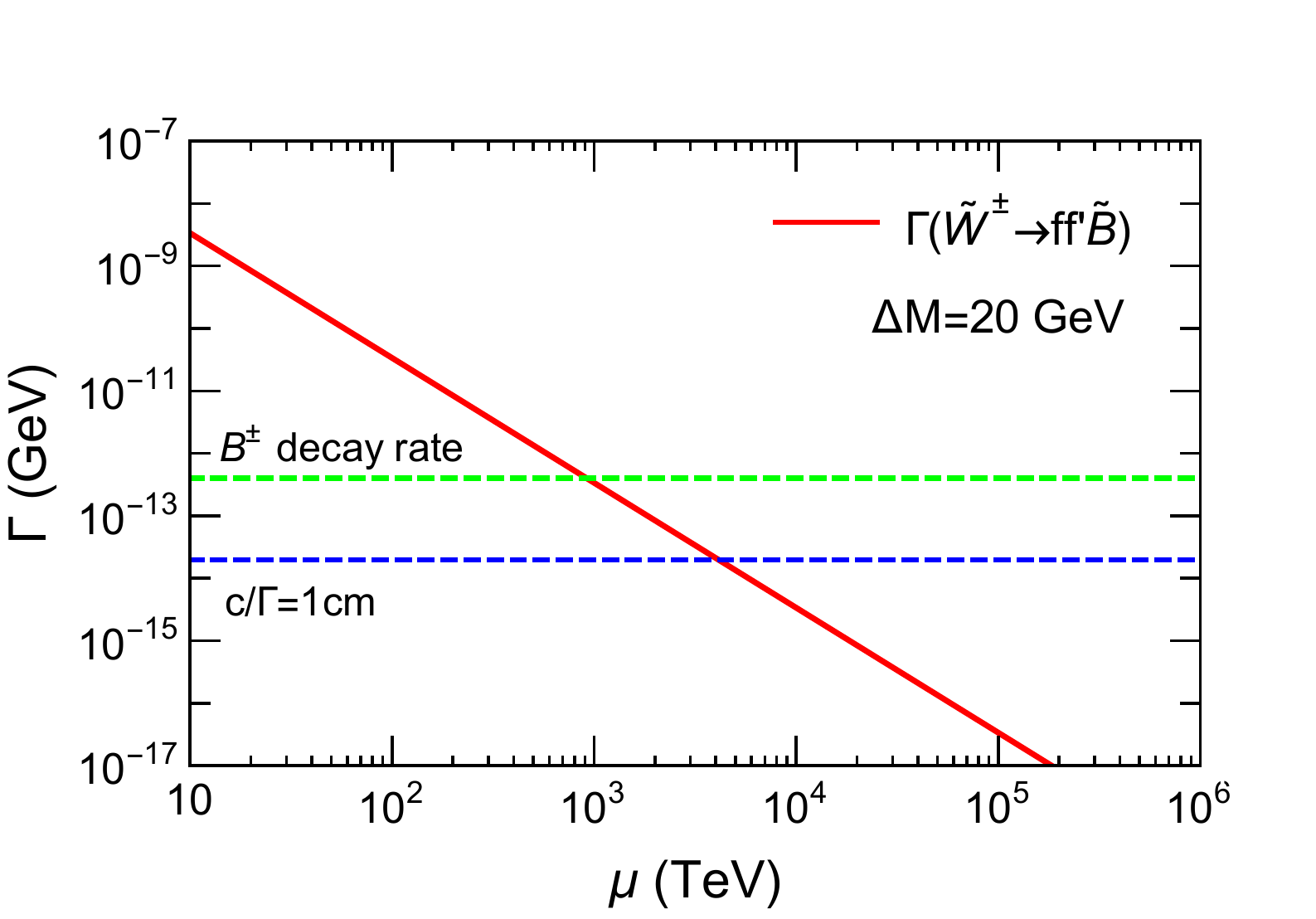}\label{fig:GamMu_Wpm-ffB_20}} 
\caption{ The three-body decay rate of the charged wino as a function of $|\mu|$ for two values of  $\Delta M = 50$~GeV and $20$~GeV, left and right panel, respectively.
The other parameters are fixed at: $M_2 = M_1 + \Delta M$, $M_1 = 500$ GeV, $\tan \beta = 2$.
\label{fig:GamMu_W_3B} }
\end{center}
\end{figure}

We show the $\twpm$ decay rate, $\Gamma(\twpm \to f \bar f' \tilde B) = \sum_f \Gamma_f(\twpm \to f \bar f' \tilde B)$, in figure~\ref{fig:GamMu_Wpm-ffB_50}
and \ref{fig:GamMu_Wpm-ffB_20} for $\Delta M = 50$ and $20$~GeV, respectively.
It is clear that the charged wino can be long lived for $|\mu| \gsim {\cal O}(10^{3-4})$~TeV depending on the mass splitting $\Delta M$.
The approximate formula for the charged wino lifetime is given by:
\beq
c \tau_{\tilde W^\pm} \simeq 1 \,{\rm cm} \cdot
\Big( \frac{\mu}{10^4\,\tev} \Big)^{2}
\Big( \frac{30\,\gev}{ \Delta M } \Big)^{3}
\Big( \frac{1}{s_{2\beta}} \Big)^2 \, .
\eeq

\begin{figure}[t!]
\begin{center}
\includegraphics[width=0.32\textwidth]{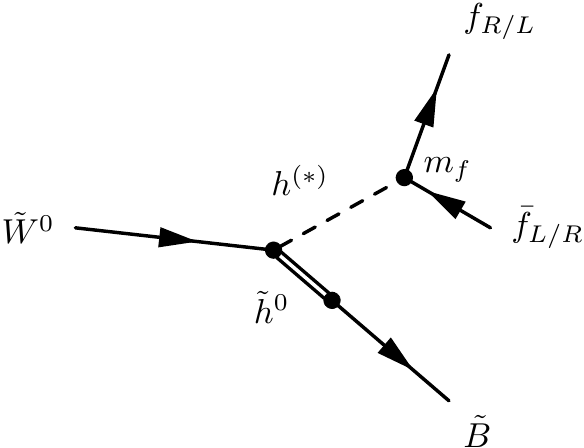}\label{fig:W-h-ffB_1}
\hspace{3mm}
\includegraphics[width=0.32\textwidth]{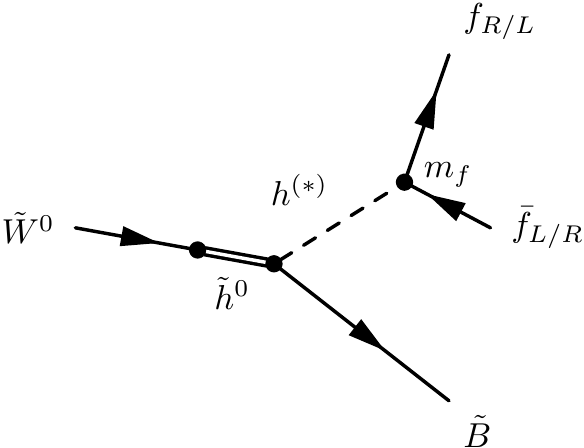}\label{fig:W-h-ffB_2}
\\
~~~~~${\cal M}^{LR}_{h} \propto m_f/\mu$
\end{center}
\begin{tabular}{cc}
\begin{minipage}{0.5\hsize}
\begin{center}
\includegraphics[width=0.65\textwidth]{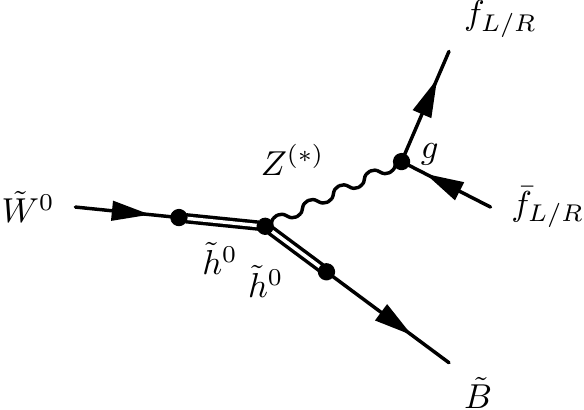}\label{fig:W-z-ffB}
\\
${\cal M}^{LL/RR}_{Z} \propto g/\mu^2$
\end{center}
\end{minipage}
\begin{minipage}{0.5\hsize}
\begin{center}
\includegraphics[width=0.65\textwidth]{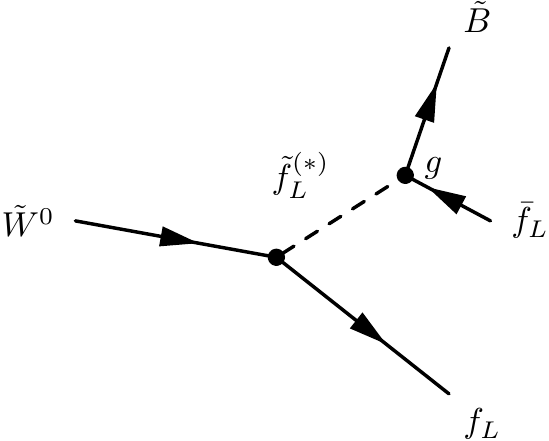}\label{fig:W-sf-ffB}
\\
${\cal M}_{\tilde f}^{LL} \propto g/m_{\tilde f_L}^2$
\end{center}
\end{minipage}
\end{tabular}
\caption{Diagrams contributing to $\twz \to f \bar f \tilde B$ three-body decay.}
\label{fig:w0_ffB}
\end{figure}

We now turn to the three-body decay of the neutral wino for $\Delta M < m_Z$.
In the previous section we have noted that the $\tilde W$-$\tilde B$-$Z$ coupling is suppressed by an additional $m_Z/\mu$ factor compared to the
$\tilde W$-$\tilde B$-$h$ coupling, and therefore the  decay mode to the Higgs dominates in the neutral wino two-body decay.
In three-body decay, however, the Higgs exchange diagram also receives a suppression,
which is proportional to the mass of the final state fermions
as shown in the upper diagrams in figure~\ref{fig:w0_ffB}.
 

There are several competing contributions depending on the value of $|\mu|$.
Firstly, there is a $Z$-exchange diagram shown in the lower-left panel of figure~\ref{fig:w0_ffB}.
The matrix element of this diagram has the $LL$ and $RR$ chirality structure for the final state fermions
and is proportional to $1/\mu^2$.
Additionally, if sfermions masses are of the same order as $|\mu|$,
the sfermion exchange diagram shown in the lower-right panel of figure~\ref{fig:w0_ffB} can provide a sizeable contribution.
The matrix element is inversely proportional to the squared mass of the left-handed sfermions, $m^2_{\tilde f_L}$.
The contribution can interfere with the $Z$-exchange diagram since the matrix element has the $LL$ structure.
These contributions become significant if $|\mu|$ or $m^2_{\tilde f_L}$ are relatively small.

Let us assume that sfermions and higgsinos have the same mass scale, $M_X$,
which is much larger than $m_Z$.
Neglecting the contribution induced from the dimension-5 operator, the total decay rate can be written as
\beq
\Gamma(\twz \to f \bar f \tilde B) = \Gamma^{(2)}(\twz \to f \bar f \tilde B) + \Gamma^{(4)} (\twz \to f \bar f \tilde B)\,,
\eeq 
where $\Gamma^{(2)} = \sum_f \hat \Gamma^h_f$
is the Higgs exchange contribution which scales as $1/M^2_X$, whilst
$\Gamma^{(4)} = \sum_f \Big(
\hat \Gamma^Z_{f} + 
\hat \Gamma^{\tilde f}_{f} + 
\hat \Gamma^{V \tilde f}_{f}
\Big)
$
is the contributions from the $Z$-exchange, sfermion exchange and their interference,
scaling as $1/M^4_X$.\footnote{
The $Z$-exchange contribution induced from the dimension-5 operator has scaling $1/M_X$ in the matrix element and interferes with the tree-level $Z$ and sfermion contributions.
We neglect this contribution in this study because the operator should be suppressed by the loop factor ${\cal O}(\alpha / 4 \pi)$,
which is seemingly smaller than the $m_f/m_W$ factor for $f = b, \tau$ in the tree-level Higgs exchange contribution.
The full calculation including the higher order $Z$-exchange contribution is however important to precisely determine the lifetime of the neutral wino,
which is beyond the scope of this paper.
}

The analytic expression for the Higgs exchange contribution, $\hat \Gamma^h_f$, is given by: 
\beqn
\hat \Gamma^h_f(\twz \to f \bar f \tilde B) 
&=& 
\frac{ \alpha^2 M_2 }{32 \pi} \frac{1}{s_W^2} \cdot
\left( \frac{m_f}{m_W} r_2^f \right)^2 \cdot G^2_{\tilde W^0 \tilde B h} \cdot \Omega_h (\mu_{\tilde B}, \mu_h)
\nonumber \\
&\stackrel{ \mu, m_A \gg m_Z}{\simeq}&
\frac{\alpha^2 M_2}{8 \pi} \frac{\stb^2}{\stw^2} \frac{m_f^2}{\mu^2} \Omega_h (\mu_{\tilde B}, \mu_h)
\nonumber \\
&\stackrel{\Delta M \ll M_2}{\simeq}&
\frac{4 \alpha^2}{15 \pi} \frac{(\Delta M)^5}{m_h^4} \frac{\stb^2}{\stw^2} \frac{m_f^2}{\mu^2} \,,
\label{eq:Gh}
\eeqn
where $\mu_h = m_h^2/M_2^2$ and $r_2^u = c_\alpha / s_\beta$ and $r_2^d = - s_\alpha / c_\beta$, both of which are reduced to $-1$
in the decoupling limit of the SUSY Higgs bosons ($m_A \gg m_Z$).
An analytic form of  $\Omega_h (x, y)$ is given in appendix~\ref{app:funcs}, eqs.~\eqref{eq:omegah}--\eqref{eq:Hxy}.
In the last step we approximate the expression assuming $\Delta M \ll M_2$.

In the limit of heavy sfermions and higgsinos, $\hat \Gamma^Z_f$, $\hat \Gamma^{\tilde f}_f$ and $\hat \Gamma^{V \tilde f}_f$ are given by
\beqn 
\label{eq:gammaZ}
\hat \Gamma^Z_f (\tilde W^0 \to f \bar f \tilde B) 
&\stackrel{\mu \gg m_Z}{\simeq}&
\frac{\alpha^2 M_2}{32 \pi} \frac{\kappa_Z^f \ctb^2}{2\stw^2} \frac{m_Z^4}{\mu^4} \Omega_+(\mu_{\tilde B},\mu_Z) \nonumber \\
&\stackrel{\Delta M \ll M_2}{\simeq}&
 \frac{\alpha^2}{20 \pi} \kappa_Z^f \frac{\ctb^2}{\stw^2} \frac{(\Delta M)^5}{\mu^4} ,
\\
\label{eq:gammaf}
\hat \Gamma^{\tilde f}_f (\tilde W^0 \to f \bar f \tilde B) 
&\stackrel{\mu,\, m_{\tilde f_L} \gg m_Z}{\simeq}&
\frac{\alpha^2 M_2}{32 \pi} \hat \kappa_f^2 
\frac{M_2^4}{m^4_{\tilde f_L}}  
\Omega_f(\mu_{\tilde B})
\nonumber \\
&\stackrel{\Delta M \ll M_2}{\simeq}&
\frac{\alpha^2}{12 \pi} \hat \kappa_f^2 \frac{(\Delta M)^5}{m^4_{\tilde f_L}} ,
\\
\hat \Gamma^{V \tilde f}_f (\tilde W^0 \to f \bar f \tilde B) 
&\stackrel{\mu,\, m_{\tilde f_L} \gg m_Z}{\simeq}&
\frac{\alpha^2 M_2  }{32 \pi} \frac{\hat \kappa_f \hat v_f \ctb}{\stw} \frac{ M_2^2}{m^2_{\tilde f_L}} 
\frac{m^2_Z}{\mu^2}
\Omega_{Vf}(\mu_{\tilde B}, \mu_Z)
\nonumber \\
&\stackrel{\Delta M \ll M_2}{\simeq}&
- \frac{\alpha^2}{15 \pi} \hat \kappa_f \hat v_f \frac{\ctb}{\stw} \frac{(\Delta M)^5}{m^2_{\tilde f_L} \mu^2} ,
\label{eq:gammaVf}
\eeqn
with $\mu_Z = m_Z^2/M_2^2$, $\hat \kappa_f = | \hat e_{L1}^f| | \hat e^f_{L2}|$ and 
$\hat v_f = 4(I_3^f - e_f s_W^2)$,
where $I^f_3$ and $e_f$ are the isospin and the electric charge of the sfermion $\tilde f$.
The $e_{Li}^f$ represents the coupling between the left-handed sfermion and the neutralino $\tilde \chi_i^0$, which can be written as
\beqn
 \bmat \hat e^f_{L1} \\ \hat e^f_{L2} \emat 
 &=& 
 \sqrt{2} \Big[ 
e_f \bmat c_W \\ s_W \emat + \frac{I_3^f - e_f s^2_W}{c_W s_W} \bmat - s_W \\ c_W \emat \Big].
\eeqn
The $\kappa_Z^f$ is given by $(2 I_3^f - 4 e_f \sw^2)^2 + (2 I_3^f)^2$, namely for each fermion we have
\beqn
u&:&~ (1 - \frac{8}{3} \sw^2)^2 + 1 \nonumber \\
d&:&~ (-1 + \frac{4}{3} \sw^2)^2 + 1 \nonumber \\
\nu&:&~ 2 \nonumber \\
e&:&~ (-1 + 4 \sw^2)^2 + 1 \, .
\eeqn
Analytic forms of $\Omega_+$, $\Omega_f$ and $\Omega_{Vf}$ are
given in appendix~\ref{app:funcs}.

\begin{figure}[t!]
\begin{center}
  \subfigure[]{\includegraphics[width=0.5\textwidth]{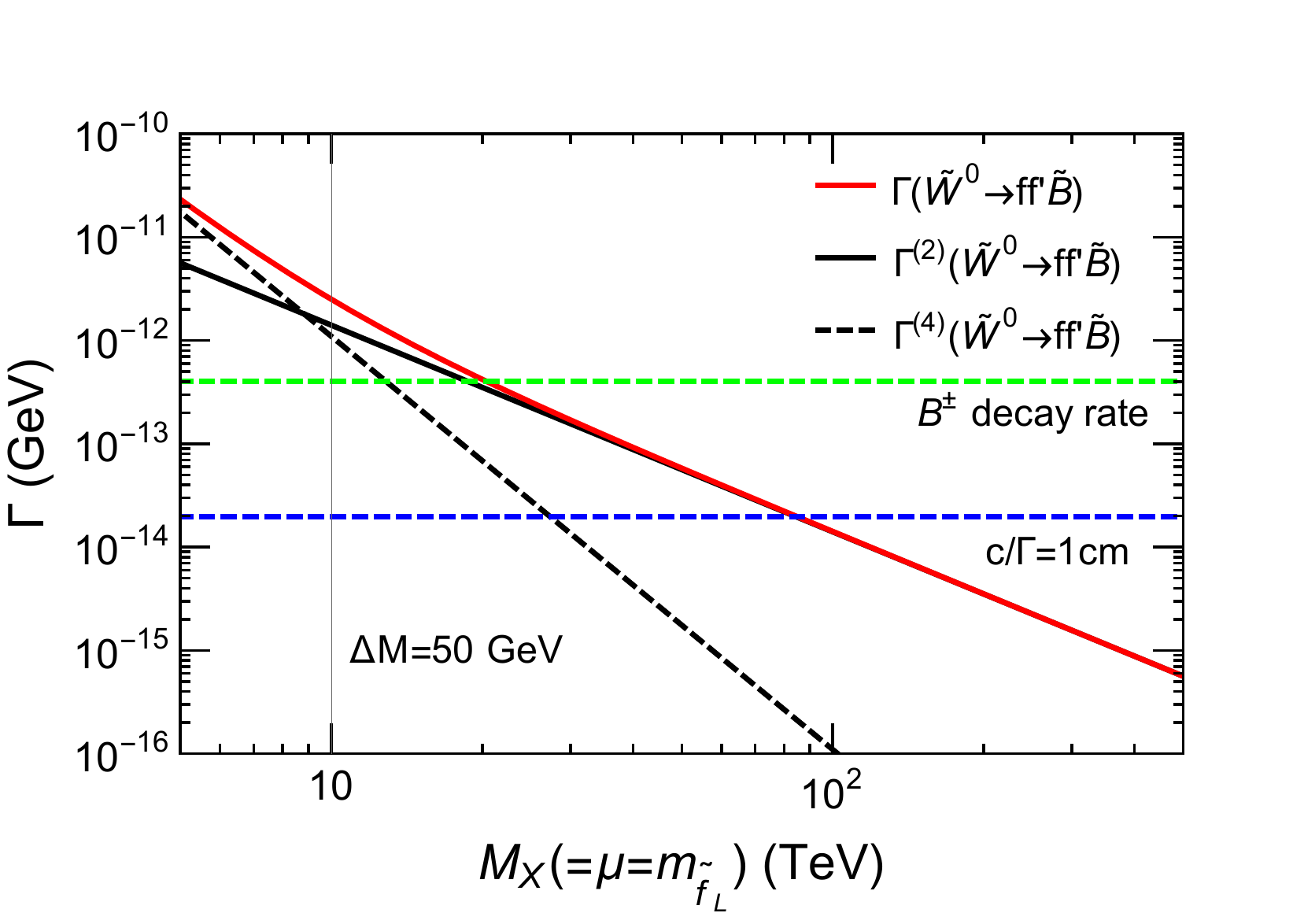}\label{fig:GamMu_W0-ffB_50}} \hspace{-4mm}
  \subfigure[]{\includegraphics[width=0.5\textwidth]{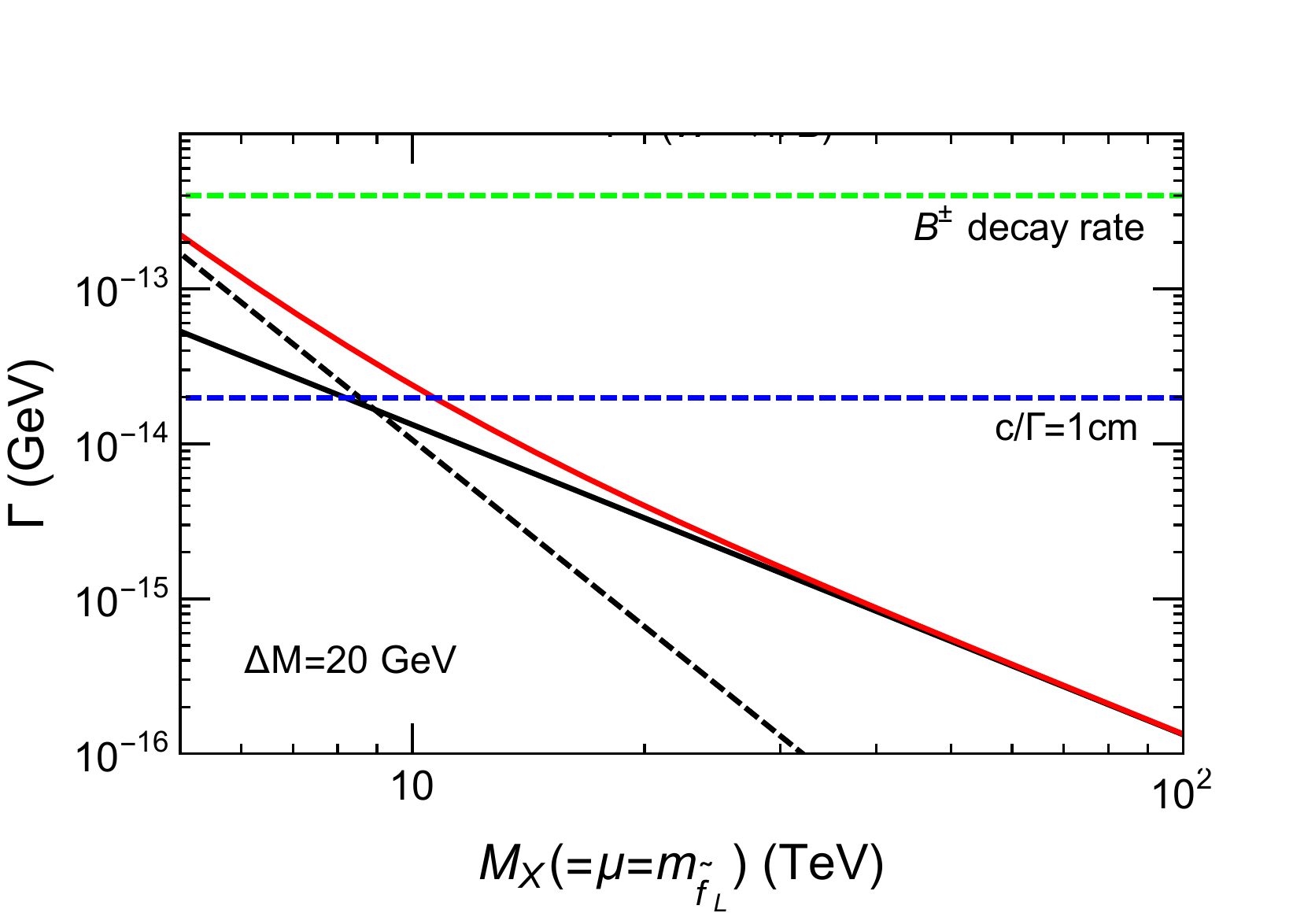}\label{fig:GamMu_W0-ffB_20}} 
\caption{ The three-body decay rate of the neutral wino as a function of $M_X (= |\mu| = m_{\tilde f_L})$ for 
two values of $\Delta M = 50$~GeV and $20$~GeV, left and right panel, respectively. The other parameters are taken at
$M_2 = M_1 + \Delta M$, $M_1 = 500$ GeV, $\tan \beta = 2$.
 \label{fig:GamMu_W0_3B} }
\end{center}
\end{figure}

The contributions $\Gamma^{(2)}(\twz \to f \bar f \tilde B)$ and $\Gamma^{(4)} (\twz \to f \bar f \tilde B)$
as well as the total decay rate $\Gamma(\twz \to f \bar f \tilde B)$ are shown in figure~\ref{fig:GamMu_W0-ffB_50} and \ref{fig:GamMu_W0-ffB_20}
for $\Delta M = 50$ and 20~GeV, respectively, as functions of $M_X$, assuming $\mu = m_{\tilde f_L} \equiv M_X$.
One can see that the neutral wino can become long lived
with $M_X \gsim 20$~TeV for the $\Delta M = 50$~GeV case 
and even for $M_X \gsim 5$~TeV for the $\Delta M = 20$~GeV case.
The contribution from the $Z$ and sfermion exchange diagrams, $\Gamma^{(4)}$, is negligible for very large $\mu$.
As  $M_X$ is decreasing, this contribution starts to be significant around $M_X \sim 20$~TeV and becomes the leading contribution 
from $M_X \lsim 8$~TeV.

In the limit of $\mu, m_{\tilde f_L}, m_A \gg m_Z$ and $\Delta M \ll M_2$, the lifetime of the neutral wino can be written approximately as
\beqn
c \tau_{\tilde W^0} &\simeq & 1 \,{\rm cm} \cdot
\Big( \frac{\mu}{10^2\,\tev} \Big)^{2}
\Big( \frac{50\,\gev}{ \Delta M } \Big)^{5}
S^{-1}\\
&\stackrel{\mu \simeq m_{\tilde f_L} > 20/\sin(2\beta)\,\mathrm{TeV}}{\simeq}& 1 \,{\rm cm} \cdot
\Big( \frac{\mu}{10^2\,\tev} \Big)^{2}
\Big( \frac{50\,\gev}{ \Delta M } \Big)^{5}
\Big( \frac{1}{s_{2\beta}} \Big)^{2} \,,\label{eq:wino3bappr}
\eeqn
with
\beq
S = 
s_{2\beta}^2 
+
c_{2\beta}^2
\Big( \frac{7.87\,{\rm TeV}}{\mu} \Big)^2
+
\sum_i \Big[
\hat \kappa_{f_i}^2 
\Big( \frac{\mu}{m_{\tilde f_{L_i}}} \Big)^2
\Big( \frac{1.56\,{\rm TeV}}{m_{\tilde f_{L_i}}} \Big)^2 
-
c_{2\beta} 
\hat \kappa_{f_i} \hat v_{f_i} 
\Big( \frac{1.53\,{\rm TeV}}{m_{\tilde f_{L_i}}} \Big)^2
\Big] \,.\label{eq:sfactor}
\eeq
For $|\mu| \simeq m_{\tilde f_L} > 20 / \sin(2\beta)$~TeV, the total decay rate is dominated\footnote{At $|\mu|=m_{\tilde f_L} = 20\, \mathrm{TeV}$ and for $\tan\beta =2$ we have $\Gamma^{(4)}/\Gamma^{(2)} = 0.195$.} by $\Gamma^{(2)}$ and
in this regime, the lifetime of the neutral wino can be approximated by eq.~\eqref{eq:wino3bappr}. Note that $\hat \kappa_{f_i}^2$can be $\mathcal{O}(1)$, table~\ref{tab:kappa}, and in case when there is an additional large hierarchy between $\mu$ and $m_{\tilde f_L}$ the third term of eq.~\eqref{eq:sfactor} might again become relevant.

\begin{table}[t]
\begin{center}
\begin{tabular}{lcccc}\toprule
 & $u$ & $d$ & $e$ & $\nu$ \\ \midrule
 $\hat \kappa_{f}^2$ & $0.16$   & $0.16$   & $1.44$  &  $1.44$ \\ 
 $\hat \kappa_f \hat v_f$  & $0.56$   & $0.68$   &  $-1.33$  &  $-2.40$  \\ 
 $\kappa_Z^f$ & $1.16$   & $1.49$   & $1.01$   & $2$ \\ \bottomrule
\end{tabular}
\end{center}
\caption{Approximate values of coefficients $\hat \kappa_{f}^2$, $\hat \kappa_f \hat v_f$ and $\kappa_Z^f$ appearing in eqs.~\eqref{eq:gammaZ}--\eqref{eq:gammaVf}.\label{tab:kappa} }
\end{table}


\subsection{Bino NLSP case ($|M_1| > |M_2|$)\label{sec:4.2}}

In this subsection we assume bino is heavier than winos and define 
$\Delta M \equiv |M_1| - |M_2| > 0$.
If $\Delta M < m_W$, only three-body bino decays are allowed.
As discussed in the previous subsection, the Higgs exchange diagram is suppressed by the mass of the final state fermions,
and the $W$-exchange diagram for  $\tilde B \to f \bar f' \tilde W^\pm$ dominates bino decay.
The partial decay rate can be written as
\beqn
\Gamma_f(\tilde B \to f \bar f' \tilde W^\pm) 
&=& 
\frac{\alpha^2 M_1 }{16 \pi} \frac{1}{\sw^2}   \cdot G_{\tilde B \tilde W^\pm W}^2 \cdot \Omega_- (\mu_{\tilde W}, \mu_W)
\nonumber \\
&\stackrel{\mu \gg m_Z}{\simeq}&
\frac{\alpha^2 M_1 }{16 \pi} \frac{\stb^2}{\sw^2} \frac{m_Z^2}{\mu^2} 
 \frac{m_W^2 \Omega_-(\mu_{\tilde W}, \mu_W)}{(M_1 - M_2)^2} 
\nonumber \\
&\stackrel{\Delta M \ll M_1}{\simeq}&
\frac{4 \alpha^2 }{15 \pi} \frac{ (\Delta M)^3 }{\mu^2} \frac{\stb^2}{\stw^2} \,,
\eeqn
where  $\mu_W \equiv m_W^2 / M_1^2$ and we approximate the expression assuming $\Delta M \ll M_1$  on top of $\mu \gg m_Z$ in the final step. 
An analytic formula for $\Omega_- (\mu_{\tilde W}, \mu_W)$ is provided in appendix~\ref{app:funcs}, eqs.~\eqref{eq:omegaapp}--\eqref{eq:ellxy}.


\begin{figure}[t!]
\begin{center}
  \subfigure[]{\includegraphics[width=0.5\textwidth]{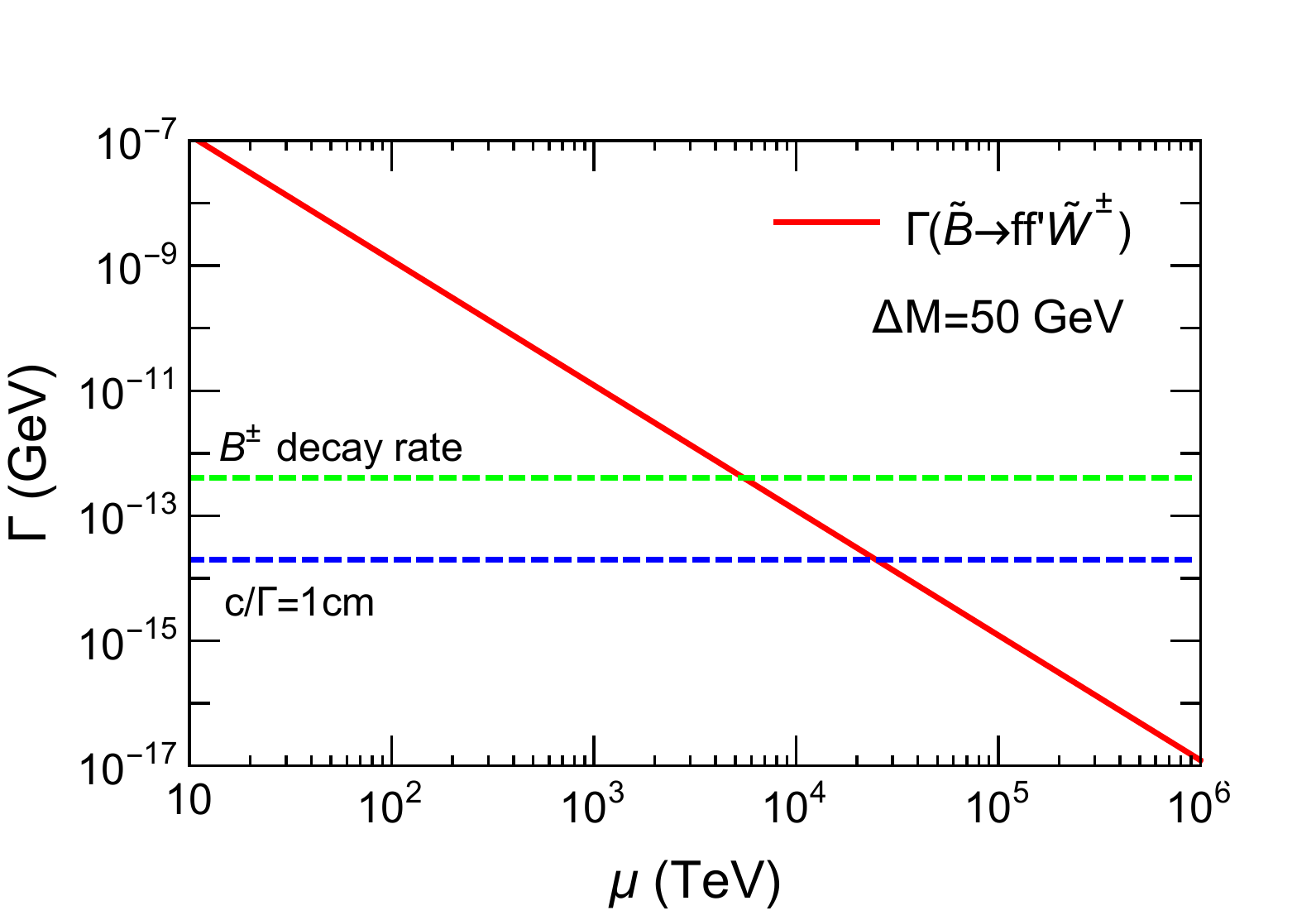}\label{fig:GamMu_B-ffW_50}} \hspace{-4mm}
  \subfigure[]{\includegraphics[width=0.5\textwidth]{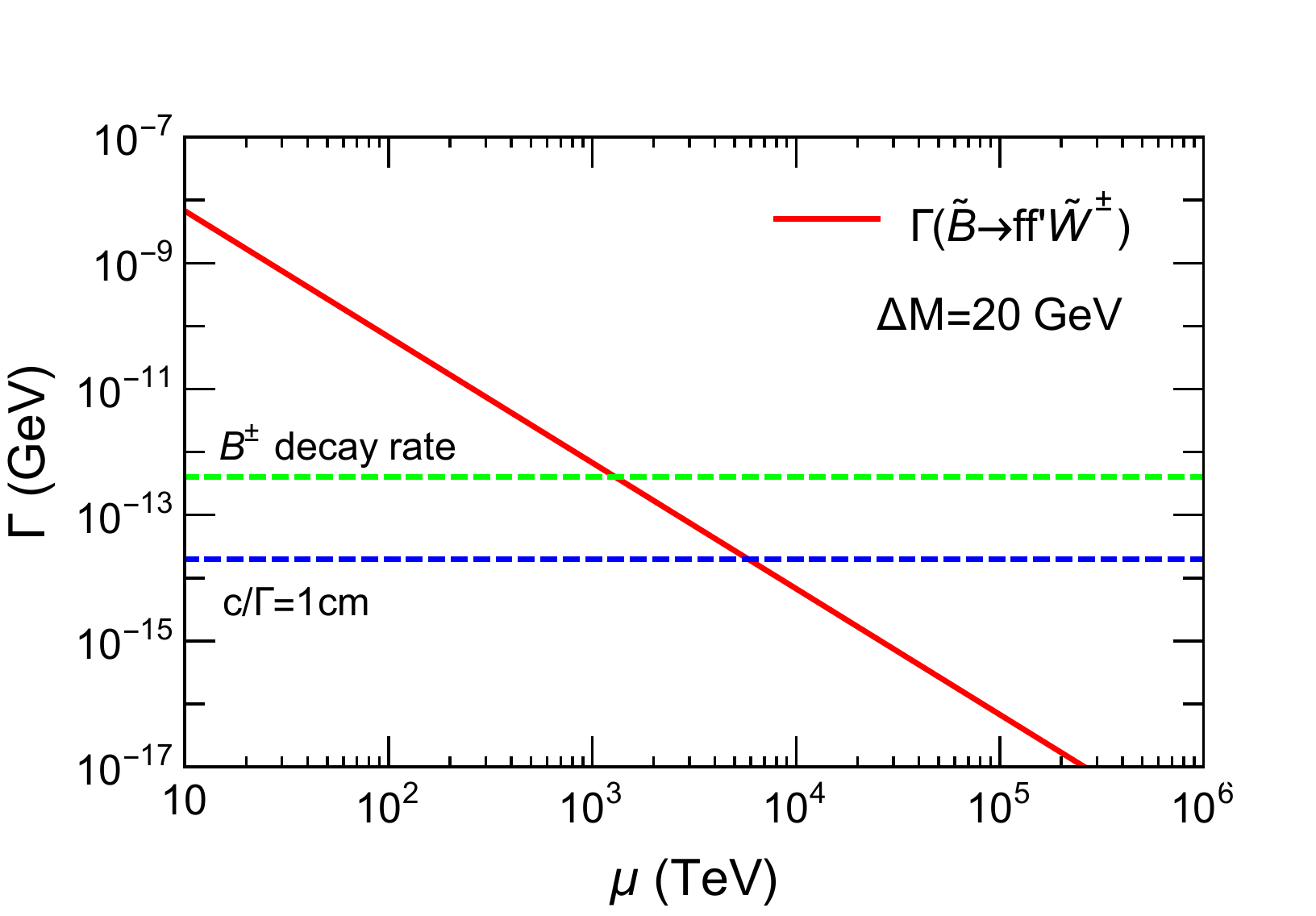}\label{fig:GamMu_B-ffW_20}} 
\caption{ The three-body decay rate of the bino as a function of $|\mu|$  for two values of  $\Delta M = 50$~GeV and $20$~GeV, left and right panel, respectively.
The other parameters are taken at
$M_1 = M_2 + \Delta M$, $M_2 = 500$~GeV, $\tan \beta = 2$.
 \label{fig:GamMu_B_3B} }
\end{center}
\end{figure}


In figures~\ref{fig:GamMu_B-ffW_50} and \ref{fig:GamMu_B-ffW_20} we show the bino decay rate
$\Gamma(\tilde B \to f f' \twpm) = \sum_f \Gamma_f(\tilde B \to f f' \twpm)$ as a function of $|\mu|$
for $\Delta M = 50$ and 20 GeV, respectively.
As can be seen, the bino becomes long lived for $|\mu| > {\cal O}(10^{3-4})$~TeV depending on $\Delta M$.
The approximate formula for the bino lifetime is given by
\beq
c \tau_{\tilde B} \simeq 0.5 \,{\rm cm} \cdot
\Big( \frac{\mu}{10^4\,\tev} \Big)^{2}
\Big( \frac{30\,\gev}{ \Delta M } \Big)^{3}
\Big( \frac{1}{s_{2\beta}} \Big)^{2} \,.
\eeq
Note that if the decay is mainly mediated by the Higgs boson exchange, the final state will be dominated by (possibly displaced) $b$ or $\tau$ jets.


\section{Large mass splitting between gauginos and higgsinos}\label{sec:splitting}

In the previous sections, we found that wino and bino can be long-lived for $|\mu| \gsim {\cal O}(10^{5-6})$ TeV 
when two-body decays are allowed, and $|\mu| \gsim {\cal O}(10^{1-4})$ TeV for three-body decays.
On the other hand, the wino and bino masses should be less than ${\cal O}(1)$ TeV to be produced at colliders. 
The long-lived wino and bino in this scenario can be observed
only if an enormous mass splitting between gauginos and higgsinos is realised. 

The higgsino mass, $\mu$, is the only dimensionfull parameter in the MSSM Lagrangian.
Since it is a supersymmetric mass term, its origin can be different from  the soft masses of gauginos and scalars.
If one does not ask about the origin of the $\mu$ term
and does accept the fine tuning in the electroweak symmetry breaking, 
phenomenologically there is no problem in taking $\mu$ at any value, as long as phenomenologically correct EWSB is achieved. 
 
If $\mu$ is very large, the low energy effective lagrangian is obtained by integrating out the heavy higgsino fields.
In doing so, the electroweak gauginos receive the threshold correction from the heavy higgsinos as \cite{Randall:1998uk, Giudice:1998xp}
\beq
\Delta M_1^{\tilde h} = \frac{\alpha_1}{4 \pi} \frac{3}{5} L\,,~~~~
\Delta M_2^{\tilde h} = \frac{\alpha_2}{4 \pi} L\,,
\label{eq:threshold}
\eeq
with
\beq
L \equiv \frac{\mu m_A^2 s_{2 \beta} }{|\mu|^2 - m_A^2} \ln \big( \frac{|\mu^2|}{m_A^2} \big)\,,
\eeq
where $m_A$ is the CP-odd Higgs mass.
If $m_A$ and $\mu$ are of the same order and $|\mu| \gsim {\cal O}(10^{3})$~TeV, these threshold corrections exceed 
${\cal O}(10)$ TeV and the Wino and Bino become out of the collider reach.
In order for these corrections to be small and realise the large mass splitting between heavy higgsinos and light gauginos,
one needs $m_A \ll |\mu|$.
Since $m_A$ is related to $\mu$ by 
\beq
m^2_A = 2 |\mu|^2 + m^2_{H_u} + m^2_{H_d}\,,
\eeq
some unnatural cancellation may be required
to accommodate observable long-lived winos and binos for $|\mu| \gsim {\cal O}(10^3)$ TeV. 

Another possibility is to attribute the origin of $\mu$ to the SUSY breaking.
For example, in gravity mediation, $\mu$ can be generated as \cite{Giudice:1988yz, Inoue:1991rk}
\beq
\mu = c \, m_{3/2},
\eeq
where $c$ is a dimensionless coefficient. 
In this scenario the gaugino masses receive the anomaly mediated contribution \cite{Randall:1998uk, Giudice:1998xp},
\beq
\Delta M_1^{\rm AM} = \frac{\alpha_1}{4 \pi} \frac{33}{5} m_{3/2}\,,~~~~~
\Delta M_2^{\rm AM} = \frac{\alpha_1}{4 \pi} m_{3/2}\,,~~~~~
\Delta M_3^{\rm AM} = - \frac{\alpha_1}{4 \pi} 3  m_{3/2}\,,
\label{eq:AM}
\eeq
as well as the higgsino threshold correction of eq.~(\ref{eq:threshold}).
The size of these correction exceeds ${\cal O}(10)$~TeV for $|\mu| \gsim {\cal O}(10^3)$~TeV.
In this scenario, some unnatural cancellation among these contributions may be required 
for the long-lived bino and winos within the collider reach for $|\mu| \gsim {\cal O}(10^3)$~TeV. 

From these considerations one may find the case of the wino NLSP with $|M_2| - |M_1| < m_Z$ particularly interesting
because the neutral wino can become long-lived for $|\mu| \gsim {\cal O}(10^{1-2})$~TeV.\footnote{
This mass spectrum of bino and wino is also motivated by a recent scan of the 10-dimensional parameter space of phenomenological MSSM \cite{deVries:2015hva}.}
With $|\mu| \sim {\cal O}(10^{1-2})$ TeV, the contributions from the higgsino threshold corrections and the anomaly mediation 
are of the order of ${\cal O}(10^{2-3})$ GeV and a large variety of the wino and bino spectra,
including compressed spectrum for the three-body decays, can be naturally achieved 
by the interplay between these contributions 
\cite{Bhattacherjee:2012ed, Hall:2012zp, Harigaya:2014dwa, Bagnaschi:2014rsa}.
In order to precisely predict the lifetime of the neutral wino in this regime, the calculation of the $Z$-exchange diagram  
induced by the dimension-5 operator may be necessary.
We leave this task for future work.


\section{Collider signatures}\label{sec:collider}

In supersymmetry with heavy scalars and higgsinos,
the bino cross section is largely suppressed both at hadron and $e^+ e^-$ colliders.
The binos may nevertheless be produced from decays of gluinos at hadron colliders.
As shown in sections \ref{sec:2body} and \ref{sec:3body}, the lifetime of bino can become larger than ${\cal O}(1)$~cm 
for $|\mu| \gsim {\cal O}(10^6)$~TeV if the two-body decay is allowed 
and $|\mu| \gsim {\cal O}(10^4)$~TeV otherwise.  

If squarks are heavy the gluinos may also become long-lived.  The gluino lifetime is given by~\cite{ArkaniHamed:2004fb}:
\beq
c \tau_{\tilde g} = {\cal O}(1~{\rm cm}) \cdot \Big( \frac{1~{\rm TeV}}{m_{\tilde g}} \Big)^{5} \Big( \frac{m_{\tilde q}}{10^3~{\rm TeV}} \Big)^{4} .
\eeq
Therefore, if the squark masses and $|\mu|$ are of similar order,
the long-lived binos are produced from long-lived gluinos. 
In order to have the gluino decay well inside the detector, $|\mu|$ and the squark mass cannot be too large. A number of searches have 
been performed by the LHC experiments looking for metastable gluinos that hadronizes into a colorless $R$-hadron. These include cases when the $R$-hadron
escapes a detector before decaying~\cite{Aad:2012pra,Chatrchyan:2013oca}, the $R$-hadron is stopped in a detector~\cite{Aad:2013gva,Khachatryan:2015jha,Aad:2015qfa}, or it decays inside the detector~\cite{ATLAS-CONF-2014-037}.

Bino can become long-lived already with  $|\mu|\sim {\cal O}(10^4)$~TeV. 
In this case the bino predominantly decays into an off-shell $W$ and a charged wino.
The charged wino subsequently decays into an off-shell $W$ and the neutral wino, with potentially long lifetime 
\cite{Ibe:2006de, Buckley:2009kv}. 
The charged wino lifetime in this region is given by 
\beq
c \tau_{\tilde W^\pm} \simeq 5\,{\rm cm} \cdot 
\Big( \frac{160\,{\rm MeV}}{\Delta m_{\tilde W}} \Big)^{3}
\Big( 1 - \frac{m^2_\pi}{\Delta m^2_{\tilde W}} \Big)^{-1/2}\,,
\eeq
where the mass splitting $\Delta m_{\tilde W} \equiv m_{\tilde W^\pm} - m_{\tilde W^0}$ can be written as
\beq
\Delta m_{\tilde W}^{\rm tree} \sim \frac{1}{4} {\rm sign}(M_1 M_2) \frac{m_Z^4 s^2_{2W} s^2_{2\beta}}{\mu^2 |M_2 - M_1|}\,,
\eeq
at tree-level and receives a radiative correction, $\Delta m_{\tilde W}^{\rm rad} \sim 160$~MeV \cite{Ibe:2012sx}.
The long-lived charged winos are searched for by looking for the disappearing track signature~\cite{Aad:2013yna, CMS:2014gxa} 
or measurements of ionisation energy loss in a pixel detector~\cite{Aad:2015qfa}.
If the long-lived binos carry the charged winos in the middle of the tracking system (50--100~mm),
the signal can be seen as an appearing-and-disappearing track signature in the detector.

The long-lived wino can be produced either directly at hadron and $e^+ e^-$ colliders or
indirectly from prompt/non-prompt decays of gluinos at hadron colliders.
The charged wino predominantly decays into an on- or off-shell $W$ and the LSP bino.
If the decay products of $W^{\pm (*)}$ are reconstructed, the signal may be detected 
as a kinked tracks in the leptonic $W^{(*)}$ channel or a displaced dijet with a charged track pointing to the
secondary vertex in the hadronic $W^{(*)}$ channel.

The neutral wino predominantly decays into a $h^{(*)}$ and a bino in most cases.
The signal should be detected as the displaced jets/dijet signature~\cite{CMS:2014wda,Aad:2015uaa,Aad:2015rba}. It is worth noting that the Higgs mediation can be pinned down by confirming that the displaced jets originate from $b$-quarks and $\tau$-leptons~\cite{ATLAS-CONF-2014-046,Aad:2014rga,Chatrchyan:2012jua,Chatrchyan:2012zz}, as opposed to the $Z$ mediation case, where light jets would dominate the final state. 
In the region where the neutral wino is long-lived with $|\mu| \sim {\cal O}(10^{1-2})$ TeV,
the two-body decay is kinematically forbidden and the $\twz$ decays into 
a bino and an off-shell $Z$, $h$ or left-handed sfermions with a small bino-wino mass splitting. 
Because the decay products in these off-shell decays are soft,
the $e^+ e^-$ may offer the best opportunity to detect the long-lived neutral wino in this parameter region.
At hadron colliders, the search for the displaced $Z^{(*)} \to \ell \ell$ may also be promising~\cite{CMS:2014hka,CMS-PAS-EXO-14-012,Aad:2015rba}.


\section{Conclusions\label{sec:conclusion}}

We investigated a possibility of having long-lived binos and winos in SUSY models with heavy scalars and higgsinos.
In the parameter region of interest 
the $SU(2)_L$ and $U(1)_Y$ gaugino sectors
are decoupled from each other
with very small mixings proportional to $v/\mu$.
In this region, the heavier of bino and wino practically does not interact  
with the lighter one and its lifetime becomes relevant for collider experiments, $c \tau \gsim {\cal O}(1)$ cm.
 
We revisited the decays of bino and winos  
and found simple formulae for the decay rates and lifetimes, which are valid when
the scalars and the higgsinos are much heavier than the gauginos.
We have found that the long-lived bino and wino emerge
when $|\mu| \gsim {\cal O}(10^{5-6})$~TeV if the two-body decay mode is open
and $|\mu| \gsim {\cal O}(10^{3-4})$~TeV otherwise.
One exception is the case with $0 < |M_2| - |M_1| < m_Z$, where
the lifetime of the neutral wino becomes ${\cal O}(1)$ cm for $|\mu| \sim {\cal O}(10^{1-2})$ TeV,
depending on the wino-bino mass splitting,
because the Higgs exchange diagram is suppressed by the mass of the final state fermions. We compared our results to \texttt{SDecay} finding a good agreement for the range
of parameters permitted by \texttt{SDecay}.

We briefly discussed how the large mass splitting between gauginos and higgsinos 
can be achieved.
If the origin of $\mu$ is independent of the SUSY breaking, 
the large mass splitting can be realised relatively easily, 
although the threshold correction from the heavy higgsinos to the gaugino masses needs to be suppressed.
On the other hand, if the $\mu$ is linked to the SUSY breaking and in particular $|\mu| \sim m_{3/2}$,
the contributions from the higgsino threshold correction and the anomaly mediation 
become significant.  
However the large splitting is still possible if one arranges the cancellation between these contributions.
 
We also discussed a possible collider signature for the long-lived bino and wino in this scenario.
The production of bino is only possible from the decay of gluinos,
although gluinos tend also to be long-lived for $m_{\tilde q} \gsim {\cal O}(10^3)$~TeV.     
For the long-lived bino NLSP case, the charged wino may also be long-lived, because of the very small mass splitting within the wino multiplet.
In $\tilde B \to W^{\pm (*)} \twmp$ decay,
the bino may carry the long-lived charged winos into the middle of the tracking system
and the signal could be seen as an appearing-and-disappearing track signature.
  
For the wino NLSP case, the production of winos is possible either directly or indirectly from the gluino decay.
The long-lived charged wino decaying to $W^{\pm (*)}$ and $\tilde B$ can be detected  
as a kinked-track signature from the leptonic decay of $W^{(*)}$. 
The hadronic $W^{(*)}$ mode may also be seen as events with a
displaced dijet and a track pointing to the secondary vertex. 
Detecting the long-lived neutral wino with $|\mu| \sim {\cal O}(10^{1-2})$~TeV may be challenging because
the decay products will be soft due to the small wino-bino mass splitting.
In this case, $e^+ e^-$ collider may be ideal to detect the long-lived neutral wino.
Otherwise a displaced off-shell $Z^* \to \ell \ell$ decay may be promising even at hadron colliders.

\acknowledgments
K.S. thanks Satoshi Shirai for valuable discussion.
K.S. is supported in part by
the London Centre for Terauniverse Studies (LCTS), using funding from
the European Research Council 
via the Advanced Investigator Grant 267352. K.R. has been supported by the MINECO (Spain) under contract FPA2013-44773-P; 
Consolider-Ingenio CPAN CSD2007-00042; the Spanish MINECO Centro de excelencia Severo Ochoa Program under grant SEV-2012-0249; and by JAE-Doc program. 

\appendix

\section{Auxiliary functions \label{app:funcs}}

In this appendix we summarize the analytic expressions for functions used in the calculation of decay widths.
\beq f_\pm(x, y) = \sqrt{\lambda(x,y)} \eta_\pm (x,y)\,, \eeq
\beq \lambda(x,y) = 1 + x^2 + y^2 -2x -2y -2xy\,, \eeq
\beq \eta_\pm(x,y)  = (1 + x - y) + \frac{(1-x + y)(1 - x  - y)}{y} \pm 6 \sqrt{x}\,, \eeq
\beq f_h(x,y) =  ( 1 + x - y + 2 \sqrt{x} )\sqrt{\lambda(x,y)}\, , \eeq

\beq \Omega_{\pm}(x,y) = 2F(x,y) \pm G(x,y)\,,  \label{eq:omegaapp} \eeq
\beqn 
F(x,y) &=& \frac{x - 1}{6 y}[ \om(x,y) + y(5+ 5x - 7y) ] 
\nonumber \\ &&
- \frac{y}{2} \{ (1 + x - y) \ln x + [ \om(x,y) + 2 x ] {\cal L}(x,y)  \}\,,
\eeqn
\beq
G(x,y) = \sqrt{x} \big[    
4(x-1) + (1 + x - 2y) \ln x + \{ \om(x,y) - y(1 + x - y) \} {\cal L}(x,y)
\big]\,,
\eeq
\beq \om(x,y) = 1 - 2x - 2y + (y - x)^2, \eeq
\beq \label{eq:ellxy}
{\cal L}(x,y) = \frac{2}{\sqrt{-\om(x,y)}} \Big[ 
\arctan \Big( \frac{-1 + x - y}{\sqrt{-\om(x,y)}} \Big)
-
\arctan \Big( \frac{1 - x - y}{\sqrt{-\om(x,y)}} \Big)
\Big]\,,
\eeq

\beqn \label{eq:omegah}
\Omega_h(x,y) = H(x,y) + G(x,y),
\eeqn
\beqn \label{eq:Hxy}
H(x,y) &=& \frac{1}{2}(1-x)(6y -5 -5x) + \frac{1}{2}(1 - 4y - 4xy + 3y^2 + x^2) \ln x
\nonumber \\
&& + \frac{1}{2}( -5x^2 y - 3y^3 + 7y^2 + 1 -x^2 -x + x^3 - 5y + 7 x y^2 -2xy ){\cal L}(x,y)\,, \nonumber \\
\eeqn

\beqn \label{eq:omegaf}
\Omega_f(x) &=& \frac{1}{6} (1 - 8 x + 8 x^3 - x^4 - 12 x^2 \ln x ) \nonumber \\
&&+ \frac{\sqrt{x}}{6} \big[ 1 + x(9 - 9x -x^2) + 6 x (1+x) \ln x \big]\,,
\eeqn
\beqn
\Omega_V(x) &=& \frac{1}{36} \Big(-5 - 9 x + 9 x^2 + 
    5 x^3 + (3 - 9 x - 18 x^2) \ln x \Big) 
    \nonumber \\
    && - \frac{(x-1) \sqrt{4x - 1}}{6} 
     \Big[ \arctan \Big( \frac{1}{\sqrt{4 x - 1}} \Big) +  \arctan \Big( \frac{1 - 2 x}{\sqrt{4 x - 1}} \Big) \Big]\,  ,
\eeqn

\beqn \label{eq:omegaVf}
&& \Omega_{Vf}(x,y) = (1 + \sqrt{x}) \Big[ \nonumber \\
&&  
        \qquad\qquad \phantom{+}\big[ 1 - 3 x - 3 x^2 + x^3 - 3 (1 + x) y^2 + 2 y^3 \big] \frac{\ln x}{12}  
        \nonumber \\ && 
         \qquad\qquad +   \big[ 5 + 5 x^2 - 12 (y - 1) y + 2 x (1 + 6 y) \big] \frac{x - 1}{36} \nonumber \\
        && \qquad\qquad+  \big[ 1 + x^2 + x (y -2) + y - 2 y^2 \big] \big[ x^2 + (y -1)^2 - 2 x (1 + y) \big] 
        \nonumber \\ && 
        \hspace*{6cm}\times \frac{{\cal M}(x,y)}{6 \sqrt{ -x^2 - (y -1)^2 + 2 x (1 + y)}}  \Big]\,,  
\eeqn

\beqn
{\cal M}(x,y) &=& 
     \arctan \Big[ \frac{ -1 + x + y}{ \sqrt{ -x^2 - (y -1)^2 + 2 x (1 + y)} } \Big] \nonumber \\
     && - \arctan \Big[ \frac{ 1 - x + y}{ \sqrt{ -x^2 - (y -1)^2 + 2 x (1 + y)} } \Big]\,. 
\eeqn

\section{Comparison with \texttt{SDecay} \label{sec:comparison}}

\begin{figure}[!t]
\begin{center}
\subfigure[]{\includegraphics[width=0.49\textwidth]{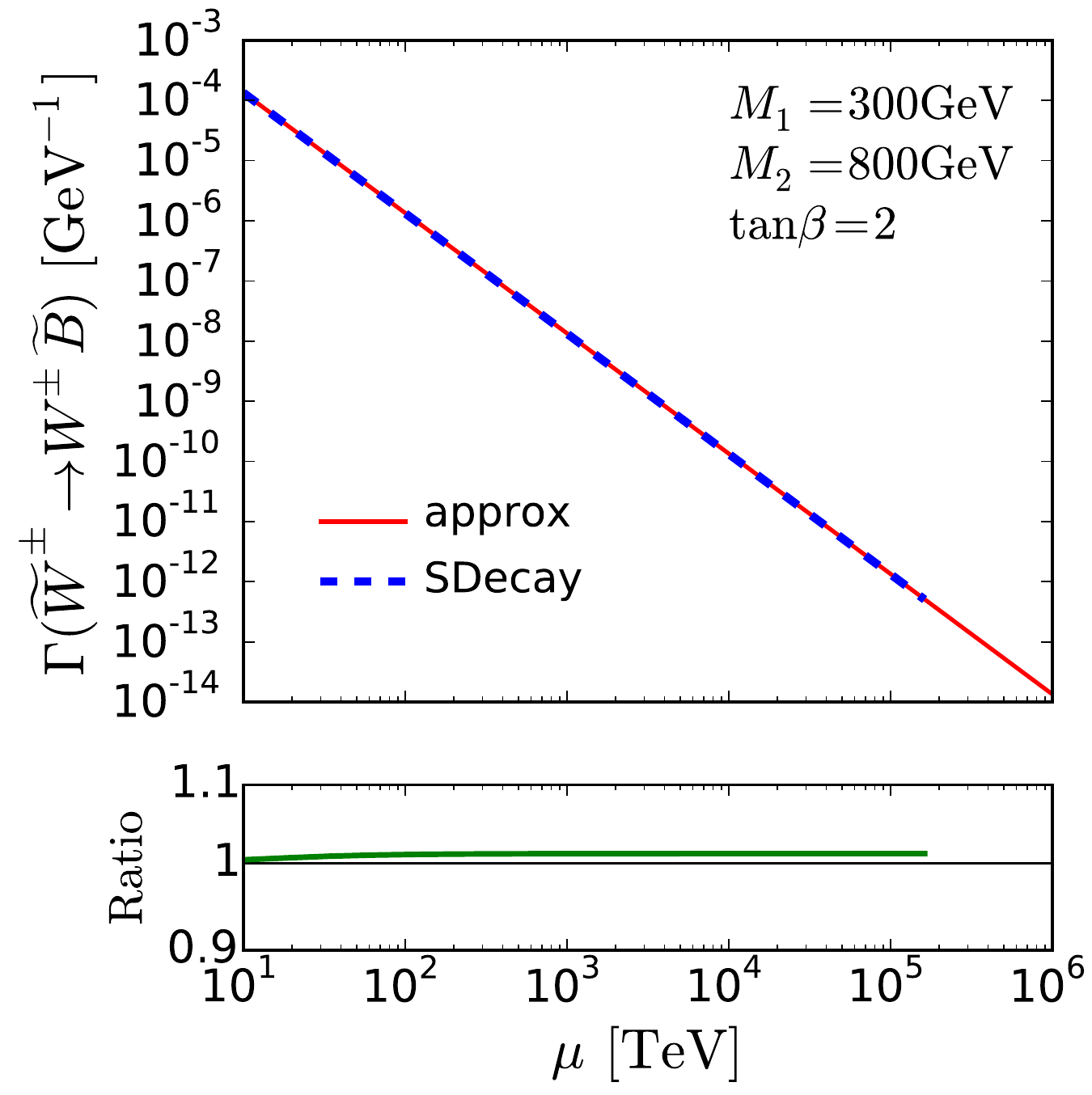}}
\subfigure[]{\includegraphics[width=0.49\textwidth]{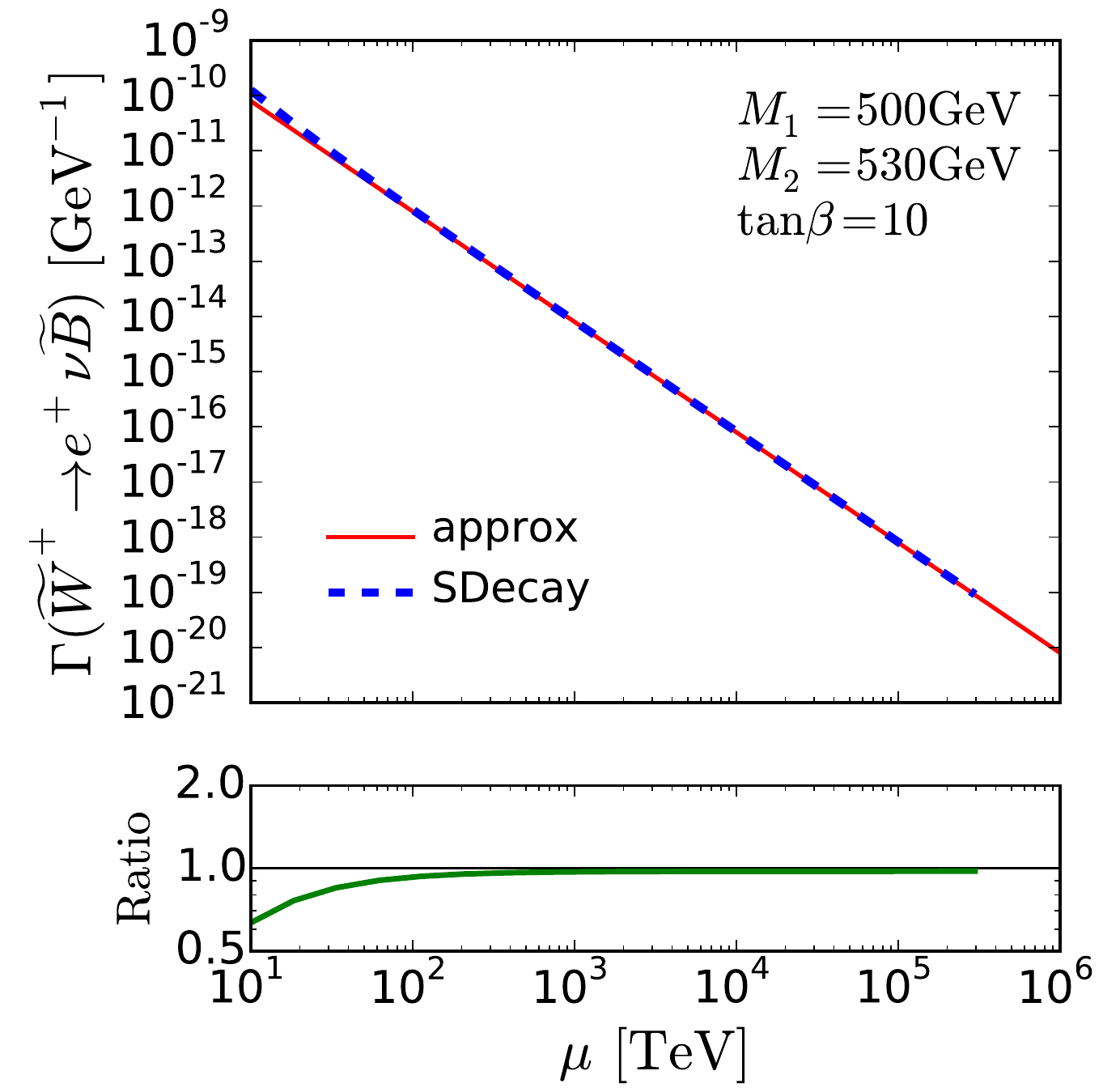}}
\subfigure[]{\includegraphics[width=0.49\textwidth]{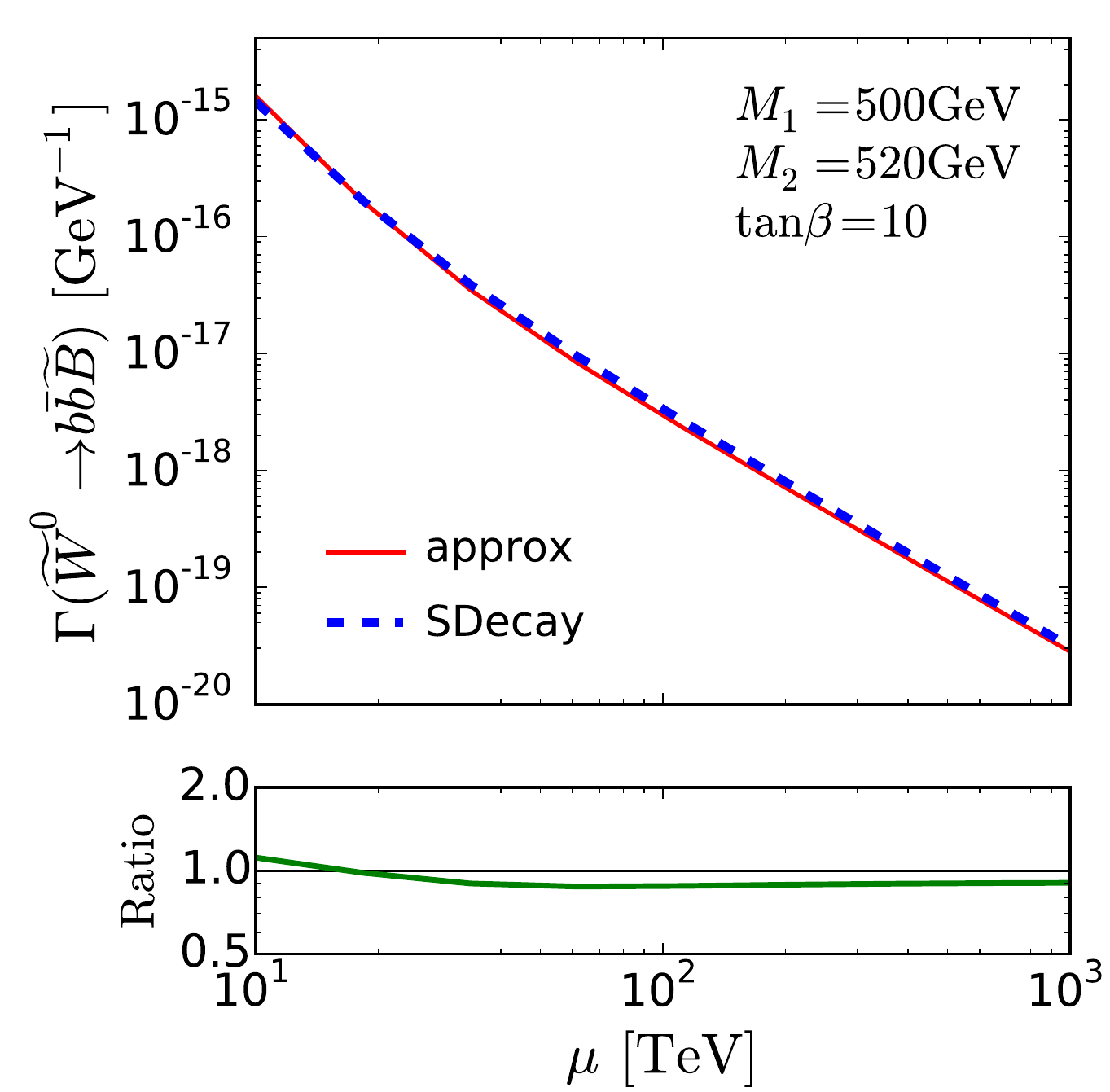}}
\caption{Comparison of partial decay widths obtained using approximated formulae (red solid line) and \texttt{SDecay}~\cite{Muhlleitner:2003vg} (blue dashed line) for different decay channels. In the lower panel of each plot the ratio of respective decay widths is shown, as defined in eq.~\eqref{eq:relative}. \label{fig:comparison}}
\end{center}
\end{figure}

In this section we compare approximated results obtained in this paper with exact calculation performed using the computer code \texttt{SDecay}~\cite{Muhlleitner:2003vg}. Figure~\ref{fig:comparison} shows the partial decay widths using both approaches for processes: $\tilde{W}^\pm \to W^\pm \tilde{B}$ (cf.\ section~\ref{sec:3.1}),  $\tilde{W}^+ \to e^+ \nu \tilde{B}$ (cf.\ section~\ref{sec:4.1}), and $\tilde{W}^0 \to b \bar{b} \tilde{B}$ 
(cf.\ section~\ref{sec:4.1}). 
For the last process we remove the squark mediated contribution by taking $m_{\tilde f_L} = 10 |\mu|$ and approximate $\Gamma = N_C (\hat \Gamma^h_b + \hat \Gamma^Z_b)$, where $N_C (= 3)$ is the colour factor. 
We use eqs.~(\ref{eq:Gh}) and (\ref{eq:gammaZ}) for 
$\hat \Gamma^h_b$ and $\hat \Gamma^Z_b$, respectively.
In addition, we show the ratio of decay widths,
\begin{equation}
 R = \frac{\Gamma^\mathrm{approx}}{\Gamma^\texttt{SDecay}}\,, \label{eq:relative}
\end{equation}
in the lower panels of each plot. In order to ensure that the spectra and couplings are the same for the purpose of comparison, we calculate the spectrum fed to \texttt{SDecay} externally to avoid corrections due to running parameters. For sufficiently large values of $\mu$ we find a very good agreement between both predictions, within $\mathcal{O}(10\%)$. 
A level of $\mathcal{O}(40\%)$ agreement can be also achieved for lower values of $\mu$. 
An important thing to note is that the calculation by \texttt{SDecay} fails for $\mu \gtrsim 10^5$~TeV, due to numerical problems. Another point to note is that because of the large separation between gaugino and stops scales one would typically run into problems with standard spectrum calculators. For these reasons using approximate formulae would be preferred over automated codes in the large $\mu$ regime.

\bibliographystyle{JHEP}
\bibliography{long_lived}

\end{document}